\documentclass[12pt,letterpaper]{article}

\usepackage{xcolor, colortbl}
\usepackage{amssymb, amsmath, amsthm}
\usepackage{graphicx}
\usepackage{natbib}
\usepackage{float}
\usepackage{setspace}
\usepackage{algorithm}
\usepackage{algorithmic}

\definecolor{lightgray}{gray}{0.9}
\newcommand{\di}{\text{d}}

\newcommand{\ba}{\mathbf{a}}
\newcommand{\bA}{\mathbf{A}}

\newcommand{\bB}{\mathbf{B}}

\newcommand{\bD}{\mathbf{D}}

\newcommand{\bbf}{\mathbf{f}}
\newcommand{\bE}{\mathbf{E}}

\newcommand{\bG}{\mathbf{G}}
\newcommand{\bH}{\mathbf{H}}
\newcommand{\bh}{\mathbf{h}}

\newcommand{\bK}{\mathbf{K}}
\newcommand{\bl}{\mathbf{l}}
\newcommand{\bL}{\mathbf{L}}

\newcommand{\bQ}{\mathbf{Q}}

\newcommand{\bS}{\mathbf{S}}
\newcommand{\bu}{\mathbf{u}}

\newcommand{\bv}{\mathbf{v}}

\newcommand{\bx}{\mathbf{x}}
\newcommand{\bX}{\mathbf{X}}
\newcommand{\by}{\mathbf{y}}
\newcommand{\bY}{\mathbf{Y}}
\newcommand{\bV}{\mathbf{V}}
\newcommand{\bW}{\mathbf{W}}
\newcommand{\bz}{\mathbf{z}}
\newcommand{\bZ}{\mathbf{Z}}

\renewcommand{\epsilon}{\varepsilon}
\renewcommand{\hat}{\widehat}
\renewcommand{\tilde}{\widetilde}
\renewcommand{\leq}{\leqslant}

\newcommand{\argmax}{\mathop{\rm argmax}}
\newcommand{\argmin}{\mathop{\rm argmin}}

\newcommand{\distn}[1]{\mathcal{#1}}

\newcommand{\Em}{\mathbb E}
\newcommand{\Pm}{\mathbb P}

\newcommand{\gvn}{\,|\,}
\newcommand{\e}{\text{e}}
\newcommand{\Var}{\text{Var}}

\newcommand{\vect}[1]{\boldsymbol #1}

\newcommand{\valpha}{\vect{\alpha}}
\newcommand{\vbeta}{\vect{\beta}}

\newcommand{\veta}{\vect{\eta}}
\newcommand{\vgamma}{\vect{\gamma}}
\newcommand{\vkappa}{\vect{\kappa}}

\newcommand{\vmu}{\vect{\mu}}
\newcommand{\vphi}{\vect{\phi}}

\newcommand{\vepsilon}{\vect{\epsilon}}

\newcommand{\vSigma}{\vect{\Sigma}}
\newcommand{\vsigma}{\vect{\sigma}}

\newcommand{\vpsi}{\vect{\psi}}
\newcommand{\vrho}{\vect{\rho}}

\newcommand{\vtheta}{\vect{\theta}}

\newcommand{\bigcdot}{\boldsymbol{\cdot}}

\DeclareMathOperator{\diag}{diag}

\begin{document}

\title{Comparing Stochastic Volatility Specifications for Large Bayesian VARs}

\author{Joshua C.C. Chan\thanks{I would like to thank Roberto Casarin, Todd Clark and William McCausland for their insightful comments and constructive suggestions. This paper has also benefited from many helpful discussions with conference and seminar participants at the 11th European Central Bank Conference on Forecasting Techniques, Ca' Foscari University of Venice, the University of Montreal and the Federal Reserve Bank of Kansas City. All remaining errors are, of course, my own.} \\
 {\small Purdue University} \\
}

\date{This version: August 2022\\ 
First version: March 2021 }

\maketitle

\onehalfspacing

\begin{abstract}

\noindent Large Bayesian vector autoregressions with various forms of stochastic volatility have become  increasingly popular in empirical macroeconomics. One main difficulty for practitioners is to choose the most suitable stochastic volatility specification for their particular application. We develop Bayesian model comparison methods---based on marginal likelihood estimators that combine conditional Monte Carlo and adaptive importance sampling---to choose among a variety of stochastic volatility specifications. The proposed methods can also be used to select an appropriate shrinkage prior on the VAR coefficients, which is a critical component for avoiding over-fitting in high-dimensional settings. Using US quarterly data of different dimensions, we find that both the Cholesky stochastic volatility and factor stochastic volatility outperform the common stochastic volatility specification. Their superior performance, however, can mostly be attributed to the more flexible priors that accommodate cross-variable shrinkage. 

\bigskip




\end{abstract}

\thispagestyle{empty}

\newpage

\section{Introduction}

Large Bayesian vector autoregressions (VARs) are now widely used for empirical macroeconomic analysis and forecasting thanks to the seminal work of \citet*{BGR10}.\footnote{The literature on using large Bayesian VARs for structural analysis and forecasting is rapidly expanding. Early applications include \citet*{CKM09}, \citet{koop13}, \citet*{KK13}, \citet{BGMR13} and \citet*{CCM15}.} Since it is well established that time-varying volatility is vitally important for small VARs, it is expected to be even more so for large systems.\footnote{For small VARs with only a few variables, papers such as \citet{clark11}, \citet*{DGG13}, \citet{CR14}, \citet{CP16} and \citet{CE18} highlight the importance of time-varying volatility for model-fit and forecasting. For large VARs with more than a dozen variables, \citet{CCM16} present evidence that the data overwhelmingly favors a VAR with stochastic volatility; \citet*{KK13}, \citet{CCM19} and \citet*{CEHK20} show that VARs with stochastic volatility forecast better than their homoskedastic counterparts, especially for density forecasts.}  Consequently, there has been a lot of recent research devoted to designing stochastic volatility specifications suitable for large systems. Prominent examples include the common stochastic volatility models \citep*{CCM16, chan20}, the Cholesky stochastic volatility models \citep{CS05,CCM19} and the factor stochastic volatility models \citep{PS99, CNS06, kastner19}. Since these stochastic volatility models are widely different and the choice among these alternatives involves important trade-offs---e.g., flexibility versus speed of estimation---one main issue facing practitioners is the lack of tools to compare these high-dimensional, non-linear and non-nested models. 

Of course, the natural Bayesian model comparison criterion is the marginal likelihood (or the marginal data density), and in principle it can be used to select among these stochastic volatility models. In practice, however, computing the marginal likelihood for high-dimensional VARs with stochastic volatility is hardly trivial due to the large number of VAR coefficients and the latent state variables (e.g., stochastic volatility and latent factors). We tackle this obstacle by developing new methods to estimate the marginal likelihood of large Bayesian VARs with a variety of stochastic volatility specifications. 

More specifically, we combine two popular variance reduction techniques, namely, conditional Monte Carlo and adaptive importance sampling, to construct our marginal likelihood estimators. We first analytically integrate out the large number of VAR coefficients---i.e., we derive an analytical expression of the likelihood unconditional on the VAR coefficients that can be evaluated quickly. In the second step, we construct an adaptive importance sampling estimator---obtained by minimizing the Kullback-Leibler divergence to the ideal zero-variance importance sampling density---to integrate out the log-volatility via Monte Carlo. By carefully combining these two ways of integration (analytical and Monte Carlo integration), we are able to efficiently evaluate the marginal likelihood of a variety of popular stochastic volatility models for large Bayesian VARs.

Compared to earlier marginal likelihood estimators for Bayesian VARs with stochastic volatility, such as \citet{CE18} and \citet{chan20}, the new method offers two main advantages. First, it analytically integrates out the large number of VAR coefficients. As such, it reduces the variance of the estimator by eliminating the portion contributed by the VAR coefficients. This reduction is expected to be substantial in large VARs. Second, earlier marginal likelihood estimators are based on \emph{local} approximations of the joint distribution of the log-volatility, such as a second-order Taylor expansion of the log target density around the mode. Although these approximations are guaranteed to approximate the target density well around the neighborhood of the point of expansion, their accuracy typically deteriorates rapidly away from the approximation point. In contrast, the new method is based on a \emph{global} approximation that incorporates information from the entire support of the target distribution. This is done by solving an optimization problem to locate the closest density to the target posterior distribution---measured by the Kullback-Leibler divergence---within a class of multivariate Gaussian distributions.

In addition to comparing different stochastic volatility specifications, the proposed method can also be used to select an appropriate shrinkage prior on the VAR coefficients. Since even small VARs have a large number of parameters, shrinkage priors are essential to avoid over-fitting in high-dimensional settings. The most prominent example of these shrinkage priors is the Minnesota prior first introduced by \citet{DLS84} and \citet{litterman86}, not long after the seminal work on VARs by \citet{Sims80}. There are now a wide range of more flexible variants \citep[see, e.g.,][]{KK93, KK97, GLP15, chan21}, and choosing among them for a particular application has become a practical issue for empirical economists. In particular, we focus on Minnesota priors with two potentially useful features: 1) allowing the overall shrinkage hyperparameters to be estimated from the data rather than fixing them at some subjectively elicited values; and 2) cross-variable shrinkage, i.e., the idea that coefficients on `other' lags should be shrunk to zero more aggressively than those on `own' lags. The proposed marginal likelihood estimators can provide a way to compare shrinkage priors with and without these features.

Through a series of Monte Carlo experiments, we demonstrate that the proposed estimators work well in practice. In particular, we show that one can correctly distinguish the three different stochastic volatility specifications: the common stochastic volatility (VAR-CSV), the Cholesky stochastic volatility (VAR-SV) and the factor stochastic volatility (VAR-FSV). In addition, the proposed marginal likelihood estimators can also be used to identify the correct number of factors in factor stochastic volatility models. 

In an empirical application using US quarterly data, we compare the three stochastic volatility specifications in fitting datasets of different dimensions (7, 15 and 30 variables). The model comparison results show that the data overwhelmingly prefers VAR-SV and VAR-FSV over the more restrictive VAR-CSV for all model dimensions. We also find strong evidence in favor of the two aforementioned features of some Minnesota priors: both cross-variable shrinkage and a data-based approach to determine the overall shrinkage strength are empirically important. In fact, when we turn off the cross-variable shrinkage in the priors of VAR-SV and VAR-FSV, they perform similarly as VAR-CSV, suggesting that the superior performance of the former two models can mostly be attributed to the more flexible priors that accommodate cross-variable shrinkage. These results thus illustrate that in high-dimensional settings, choosing a flexible shrinkage prior is as important as selecting a flexible stochastic volatility specification. Lastly, we also demonstrate how the proposed methodology can be used to improve other marginal likelihood estimators and model selection criteria such as the deviance information criterion.
 
The rest of the paper is organized as follows. We first outline in Section~\ref{s:models} the three stochastic volatility specifications designed for large BayBunesian VARs, followed by an overview of  various data-driven Minnesota priors. Then, Section~\ref{s:ML} describes the two components of the proposed marginal likelihood estimators: adaptive importance sampling and conditional Monte Carlo. The methodology is illustrated via a concrete example of estimating the marginal likelihood of the common stochastic volatility model. We then conduct a series of Monte Carlo experiments in Section~\ref{s:MC} to assess how the proposed marginal likelihood estimators perform in selecting the correct data generating process. It is followed by an empirical application in Section~\ref{s:application}, where we compare the three stochastic volatility specifications in the context of Bayesian VARs of different dimensions. Section~\ref{s:extensions} discusses various extensions of using the proposed methods to improve other marginal likelihood estimators and alternative model comparison methods. There we also consider a stochastic volatility model with an outlier component. Lastly, Section~\ref{s:conclusion} concludes and briefly discusses some future research directions.

\section{Stochastic Volatility Models for Large VARs} \label{s:models}

In this section we first describe a few recently proposed stochastic volatility specifications designed for large Bayesian VARs. We then outline a few data-driven Minnesota priors that are particularly useful for these high-dimensional models. 

\subsection{Common Stochastic Volatility}

Let $\by_t= (y_{1,t},\ldots,y_{n,t})'$ be an $n\times 1$ vector of variables that is observed over the periods $t=1,\ldots, T.$ The first specification is the common stochastic volatility model introduced in \citet*{CCM16}. The conditional mean equation is the standard reduced-form VAR with $p$ lags:
\begin{equation} \label{eq:VAR} 
	\by_t = \ba_0 + \bA_1 \by_{t-1} + \cdots + \bA_p\by_{t-p} + \vepsilon_t,
\end{equation}
where $\ba_0$ is an $n\times 1 $ vector of intercepts and $\bA_1, \ldots, \bA_p$ are all $n\times n$ coefficient matrices. To allow for heteroskedastic errors, the covariance matrix of the innovation $\vepsilon_t$ is scaled by a common, time-varying factor that 
can be interpreted as the overall macroeconomic volatility:
\begin{equation}\label{eq:csv}
	\vepsilon_t \sim\distn{N}(\mathbf{0},\e^{h_t}\vSigma).
\end{equation}
The log-volatility $h_t$ in turn follows a stationary AR(1) process:
\begin{equation}\label{eq:h}
	h_t = \phi h_{t-1} + u_t^h, \quad u_t^h\sim\distn{N}(0,\sigma^2),
\end{equation}
for $t=2,\ldots, T$, where $|\phi|<1$ and the initial condition is specified as 
$h_{1}\sim\distn{N}(0,\sigma^2/(1-\phi^2))$. Note that the unconditional mean of the AR(1) process is assumed to be zero for identification. We refer to this common stochastic volatility model as VAR-CSV.

One drawback of this volatility specification is that it appears to be restrictive. For example, all variances are scaled by a single factor and, consequently, 
they are always proportional to each other. Nevertheless, there is empirical evidence that the error variances of macroeconomic variables tend to move closely together 
\citep[see, e.g.,][]{CCM16,CES20}, and a common stochastic volatility is 
a parsimonious way to model this empirical feature. 

Estimating large VARs is in general computationally intensive because of the large number of VAR coefficients $\bA = (\ba_0, \bA_1, \ldots, \bA_p)'$. One main advantage of the common stochastic volatility specification is that---if a natural conjugate prior on $(\bA, \vSigma)$ is used---it leads to many useful analytical results that make estimation fast. In particular, there are efficient algorithms to generate the large number of VAR coefficients. Moreover, as demonstrated in \citet{chan20}, similar computational gains can be achieved for a much wider class of VARs with non-Gaussian, heteroskedastic and serially dependent innovations. Recent empirical applications using this common stochastic volatility and its variants include \citet{muntaz16}, \citet{MT17}, \citet{GH18}, \citet{Poon18}, \citet{Louzis19}, \citet{LH19}, \citet{ZN20} and \citet{Hartwig21}. 

\subsection{Cholesky Stochastic Volatility}

A more flexible way to model multivariate heteroscedasticity and time-varying covariances is to incorporate multiple stochastic volatility processes, as first proposed in \citet{CS05}. Specifically, consider the same reduced-form VAR in \eqref{eq:VAR}, but the innovation $\vepsilon_t$ has a more flexible time-varying covariance matrix modeled as follows:
\begin{equation} \label{eq:VAR-SV}
	\vepsilon_t \sim\distn{N}(\mathbf{0}, \vSigma_t), \quad \vSigma_t^{-1} = \bB_0' \bD_t^{-1}\bB_0,
\end{equation}
where $\bB_{0}$ is an $n \times n$ lower triangular matrix with ones on the diagonal and $\bD_t = \diag(\e^{h_{1,t}}, \ldots, \e^{h_{n,t}})$. The law of motion for each element of $\bh_t = (h_{1,t}, \ldots, h_{n,t})'$ is specified as an independent autoregressive process:
\begin{equation}\label{eq:hit}
	h_{i,t} = \mu_i + \phi_i(h_{i,t-1}-\mu_i) + u_{i,t}^h, \quad u_{i,t}^h \sim \distn{N}(0, \sigma_{i}^2)
\end{equation}
for $t=2,\ldots, T$, where the initial condition is specified as $h_{i,1}\sim\distn{N}(\mu_i,\sigma_i^2/(1-\phi_i^2))$. This stochastic volatility specification is sometimes called the Cholesky stochastic volatility. We refer to this stochastic volatility model as VAR-SV.

This model can be estimated using the algorithm described in \citet{DP15}. However, when the number of variables is large, the conventional way of drawing all VAR coefficients jointly becomes excessively computationally intensive. To tackle this computational problem, \citet*{CCM19} introduce a blocking scheme that makes it possible to estimate the reduced-form VAR equation by equation that drastically reduces the computational time.\footnote{As pointed out in \citet{Bognanni22}, the algorithm in \citet{CCM19} to sample the VAR coefficients equation by equation can only be viewed as an approximation, as it ignores certain integrating constants of the distributions of VAR coefficients in other equations. \citet{CCCM22} provide an alternative algorithm to draw the VAR coefficients equation by equation that has the same computational complexity of $\mathcal{O}(n^4)$.} 

In contrast to the common stochastic volatility model, VAR-SV is more flexible in that it contains $n$ stochastic volatility processes, which can accommodate more complex co-volatility patterns. But this comes at a cost of more intensive posterior computations: the complexity of estimating VAR-SV is $\mathcal{O}(n^4)$ compared to  $\mathcal{O}(n^3)$ for VAR-CSV (when a natural conjugate prior is used). Recent empirical applications using this Cholesky stochastic volatility in the context of large Bayesian VARs include \citet{BV18}, \citet{BGR18}, \citet{HF19}, \citet{CHP20}, \citet{BKL20}, \citet{KSMP20}, \citet{TZ20}, \citet{ZBZ20} and \citet{chan21}.

\subsection{Factor Stochastic Volatility}

The third stochastic volatility specification that is suitable for large systems belongs to the class of factor stochastic volatility models \citep{PS99b, CNS06, kastner19}. More specifically, consider the same reduced-form VAR in \eqref{eq:VAR}, but the innovation is instead decomposed as:
\[
	\vepsilon_t = \bL \mathbf{f}_t + \bu_t, 
\]
where $\mathbf{f}_t = (f_{1,t},\ldots, f_{r,t})'$ is a $r\times 1$ vector of latent factors and
$\bL$ is the associated $n\times r$ factor loading matrix. For identification purposes we assume that $\bL$ is a lower triangular matrix with ones on the main diagonal.\footnote{One main reason to impose these identification restrictions on $\bL$ is to avoid the so-called label switching problem, which makes it a bit more difficult to construct appropriate importance sampling densities for the log-volatility. One way to handle this issue is to first postprocess the posterior draws to sort them into the correct categories using, e.g., the approach in \citet{KS19}. Then, one can use these sorted MCMC draws to construct the importance sampling densities discussed below.} Furthermore,  to ensure one can separately identify the common and the idiosyncratic components, we adopt a sufficient condition in \citet{AR56} that requires $r \leq (n-1)/2$.\footnote{In this paper we follow the common approach of fixing the number of factors $r$ in the estimation and selecting $r$ via the marginal likelihood. An alternative approach is to allow the inclusion of potentially infinitely many factors, but to use a shrinkage prior on the factor loadings so that they are increasingly shrunk to 0 as the column index increases \citep{BD11, kastner19}. This latter approach circumvents the need to select the number of factors, but it makes the interpretation of the factors more difficult.}

The disturbances $\bu_t$ and the latent factors $\mathbf{f}_t$ are assumed to be independent at all leads and lags. Furthermore, they are jointly Gaussian:
\begin{equation} \label{eq:ft}
	\begin{pmatrix}\bu_t \\  \mathbf{f}_t \end{pmatrix} \sim\distn{N}
	\left(\begin{pmatrix} \mathbf{0}\\ \mathbf{0} \end{pmatrix},
	\begin{pmatrix} \bD_t & \mathbf{0} \\ \mathbf{0} & \bG_t \end{pmatrix}\right),
\end{equation}
where $\bD_t = \text{diag}(\e^{h_{1,t}},\ldots, \e^{h_{n,t}})$ and $ \bG_t = \text{diag}(\e^{h_{n+1,t}},\ldots, \e^{h_{n+r,t}})$ are diagonal matrices. Here the correlations among the elements of the innovation $\vepsilon_t$ are induced by the latent factors. In typical applications, a small number of factors would be sufficient to capture the time-varying covariance structure even when $n$ is large. 

Next, for each $i=1,\ldots, n+r$, the evolution of the log-volatility is modeled as:
\begin{equation}\label{eq:hit_2}
	h_{i,t} = \mu_{i} + \phi_i(h_{i,t} - \mu_i) + u_{i,t}^h, \quad u_{i,t}^h\sim\distn{N}(0,\sigma_i^2)
\end{equation}
for $t=2,\ldots, T$. The initial state is assumed to follow the stationary distribution: $h_{i,1} \sim \distn{N}(\mu_i,\sigma_i^2/(1-\phi_i^2))$. We refer to this factor stochastic volatility model as VAR-FSV.

In a sense VAR-FSV is even more flexible than VAR-SV as the former contains $n+r$ stochastic volatility processes compared to $n$ in the latter. In terms of estimation, both are more computationally intensive to estimate relative to VAR-CSV, but they can still be fitted reasonably quickly even when $n$ is large. In particular, given the latent factors VAR-FSV becomes $n$ unrelated regressions. Consequently, the $n$-equation system can be estimated equation by equation. While various factor stochastic volatility specifications are widely applied in financial applications \citep[see, e.g.,][]{AW00, JMY19, HV19, LS20, MMP20}, they are not yet commonly used in conjunction with a large VAR \citep[with the notable exception of][]{KH20}.

\subsection{Hierarchical Minnesota Priors} \label{ss:priors}

Next, we briefly describe the prior specifications for the above three stochastic volatility models; the details of the priors are given in Appendix~A. In general, we assume exactly the same priors on the common parameters across models. When this is not possible, we use similar priors so that they are comparable across models. Below we focus on the shrinkage priors on the VAR coefficients as they are critical in the current context. 

More specifically, we consider Minnesota-type priors on the VAR coefficients. This family of priors was first developed by \citet*{DLS84} and \citet{litterman86}, and a number of more flexible variants have been developed since. We adopt a version in which the overall shrinkage hyperparameters are treated as parameters to be estimated from the data, in the spirit of \citet*{GLP15}. Compared to the conventional version in which the hyperparameters are fixed at some subjective values, this data-based variant is often found to fit the data substantially better and have better out-of-sample forecast performance \citep[see, e.g.,][]{CCM15,CJZ20}. 

First, for the VAR-CSV we consider the natural conjugate prior on $(\bA, \vSigma)$ that depends on a shrinkage hyperparameter $\kappa$:
\[
	\vSigma\sim\distn{IW}(\nu_0,\bS_0), \quad (\text{vec}(\bA)\gvn\vSigma,\kappa) 
		\sim\distn{N}(\text{vec}(\bA_0), \vSigma\otimes \bV_{\bA}),
\]	
where the prior hyperparameters $\text{vec}(\bA_0)$ and $\bV_{\bA}$ are chosen in the spirit of the Minnesota prior. More specifically, since in the empirical applications we will be using data in growth rates, we set $\text{vec}(\bA_0) = \mathbf{0}$ to shrink all the coefficients to 0. For the covariance matrix $\bV_{\bA}$, it is assumed to be diagonal and it depends on a single hyperparameter $\kappa$ that controls the overall shrinkage strength (see Appendix A for details). As mentioned above, we treat $\kappa$ as an unknown parameter to be estimated. This natural conjugate prior substantially speeds up posterior simulations due to the availability of various analytical results. One major drawback of this natural conjugate prior, however, is that it cannot accommodate cross-variable shrinkage---i.e., the prior belief that coefficients on `own' lags are  on average larger than `other' lags.

Next, for the VAR-SV, we assume that the intercept and the VAR coefficients are {\em a priori} independent across equations, and each vector $\valpha_{i} = (a_{i,0},\bA_{i,1},\ldots,\bA_{i,p})'$, for $i=1,\ldots, n$, where $\bA_{i,j}$ denotes the $i$-th row of $\bA_{j}$, has a hierarchical normal prior: $\valpha_i \sim \distn{N}(\valpha_{0,i}, \bV_{\valpha_i})$. Here the prior mean $\valpha_{0,i}$ is set to be zero to shrink the VAR coefficients to zero. The prior covariance matrix $\bV_{\valpha_i}$ is assumed to be diagonal and it depends on two hyperparameters: $\kappa_1$ and $\kappa_2$. The hyperparameter $\kappa_1$ controls the overall shrinkage strength for VAR coefficients on own lags, whereas $\kappa_2$ controls those on lags of other variables. Allowing the two shrinkage parameters to be different across the two groups of VAR coefficients is empirically important. For example, \cite{CCM15} and \citet{chan21} find empirical evidence in support of cross-variable shrinkage. In addition, following recent works such as \citet{GLP15} and \citet{AMW20} that demonstrate the benefits of tuning these hyperparameters in a data-based manner relative to fixing them at specific values, we treat both $\kappa_1$ and $\kappa_2$ to be parameters to be estimated.

Let $\vbeta_i$ denote the free elements in the $i$-th row of the impact matrix $\bB_0$ for $i=2,\ldots, n$. That is, $\vbeta_i = (B_{0,i,1},\ldots, B_{0,i,i-1})'$, where $B_{0,i,j}$ is the $(i,j)$ element of $\bB_0$. We assume that $\vbeta_i$ has a hierarchical normal prior: $\vbeta_i \sim \distn{N}(\vbeta_{0,i}, \bV_{\vbeta_i})$. We set the the prior mean $\vbeta_{0,i}$ to be zero so that the free elements in the impact matrix are shrunk to zero. The prior covariance matrix $\bV_{\vbeta_i}$ is set to be diagonal, and it depends on a hyperparameter $\kappa_3$, which controls the overall shrinkage strength. We again treat $\kappa_3$ as an unknown parameter to be estimated. Allowing this shrinkage parameter to be determined by the data is found to substantially improve forecasts \citep[see, e.g.,][]{chan21}.

Finally, for the VAR-FSV we assume a similar normal prior on $\valpha_{i} = (a_{i,0},\bA_{i,1},\ldots,\bA_{i,p})'$, the intercept and the VAR coefficients in the $i$-th equation: $\valpha_i \sim \distn{N}(\valpha_{0,i}, \bV_{\valpha_i})$. Again we set $\valpha_{0,i} = \mathbf{0}$ and assume $\bV_{\valpha_i}$ to be diagonal and depends on two hyperparameters: one controls the overall shrinkage strength on coefficients on own lags and the other on coefficients on other lags. Details of these Minnesota priors and priors on other parameters are given in Appendix~A.

\section{Marginal Likelihood Estimation: Adaptive Importance Sampling and Conditional Monte Carlo} \label{s:ML}

In this section we first give an overview of the marginal likelihood and its role in Bayesian model comparison. We then describe the two components of the proposed marginal likelihood estimators: adaptive importance sampling and conditional Monte Carlo. Specifically, conditional Monte Carlo is used to integrate out the large number of VAR coefficients analytically, and adaptive importance sampling is used to bias the distributions of the remaining parameters to oversample the `important' regions of the parameter space that contribute most to the integral (and this bias is later corrected in the estimation). Finally, to illustrate the methodology, we provide the details of estimating the marginal likelihood of the VAR-CSV by combining adaptive importance sampling and conditional Monte Carlo.

\subsection{Overview of the Marginal Likelihood}

Suppose we want to compare a collection of models $\{M_{1},\ldots, M_{K} \}$, where
each model $M_k$ is formally defined by a likelihood function $p(\by\gvn \vtheta_k, M_{k})$ and a prior on the model-specific parameter vector $\vtheta_k$ denoted by $p(\vtheta_k \gvn M_k)$. A natural Bayesian model comparison criterion is the Bayes factor in favor of $M_i$ against $M_j$, defined as
\[
	\text{BF}_{ij}  = \frac{p(\by\gvn M_i)}{p(\by\gvn M_j)},
\]
where
\[
	p(\by\gvn M_{k}) = \int p(\by\gvn \vtheta_k, M_{k}) p(\vtheta_k\gvn M_{k})\di\vtheta_k
\]
is the \emph{marginal likelihood} under model $M_k$, $k=i,j.$ One can show that if the data is generated from model $M_i$, the Bayes factor would on average pick the correct model over some distinct model, say, $M_j$. To see that, we compute the expected log Bayes factor in favor of model $M_i$ with respect to the distribution $p(\by\gvn M_i)$:
\[
	\Em_{\by} \left[\log\frac{p(\by\gvn M_i)}{p(\by\gvn M_j)}\right] = \int  p(\by\gvn M_i)\, \log\frac{p(\by\gvn M_i)}{p(\by\gvn M_j)}\di \by.
\]
This expression is the Kullback-Leibler divergence from $p(\by\gvn M_i)$ to $p(\by\gvn M_j)$, and it is strictly positive unless $p(\by\gvn M_i) = p(\by\gvn M_j)$, in which case it is zero. In other words, the marginal likelihood of $M_i$ is larger than that of $M_j$ on average.

Furthermore, the Bayes factor is related to the posterior odds ratio between the two models as follows:
\[
	\frac{\Pm(M_i\gvn\by)}{\Pm(M_j\gvn\by)} = \frac{\Pm(M_i)}{\Pm(M_j)}\times \text{BF}_{ij},
\]
where $\Pm(M_i)/\Pm(M_j)$ is the prior odds ratio. It follows that if both models are equally probable \textit{a priori}, i.e., $p(M_i) = p(M_j)$, the posterior odds ratio between the two models is then equal to the Bayes factor or ratio of marginal likelihoods. In that case, if, say, $\text{BF}_{ij}=100$, then model $M_i$ is 100 times more likely than model $M_j$ given the data. For a more detailed discussion of the Bayes factor and its role in Bayesian model comparison, see \citet{koop03} or \citet{CKPT19}. From here onwards we suppress the model indicator; for example we denote the marginal likelihood and the likelihood by $p(\by)$ and $p(\by\gvn \vtheta)$, respectively. 

\subsection{Adaptive Importance Sampling}\label{ss:IS}

Next, we give an overview of an adaptive importance sampling approach
called the improved cross-entropy method for estimating the marginal likelihood. 
It is based on the idea of biasing the sampling distribution in such a
way that more `important values' are generated in the simulation. 
The sample is then weighted to correct for the use of a different distribution 
to give an unbiased estimator. In our context of estimating the marginal likelihood $p(\by)$, 
consider the following importance sampling estimator:
\begin{equation} \label{eq:ISml}
    \widehat{p(\by)}_{\rm IS} = \frac{1}{R}\sum_{r=1}^{R}
		\frac{p(\by\gvn \vect{\theta}^{(r)})p(\vect{\theta}^{(r)})}{g(\vect{\theta}^{(r)})},
\end{equation}
where $\vect{\theta}^{(1)},\ldots, \vect{\theta}^{(R)}$ are independent draws
obtained from the importance sampling density $g(\cdot)$ that dominates
$p(\by\gvn\cdot)p(\cdot)$---i.e., $g(\vtheta)=0\Rightarrow p(\by\gvn\vtheta)p(\vtheta)=0$.

The importance sampling estimator in~\eqref{eq:ISml} is unbiased and 
simulation consistent for any $g$ that dominates $p(\by\gvn\cdot)p(\cdot)$.
But its performance depends critically on the choice of $g$. Below we follow \citet{CE15} to use 
an improved version of the classic cross-entropy method 
to construct $g$ optimally. The original cross-entropy method
was developed for rare-event simulation by \citet{rubinstein97, rubinstein99}
using a multi-level procedure to obtain the optimal importance 
sampling density \citep[see also][for a book-length treatment]{rk:ce}.
 \citet{CE15} show that this optimal importance sampling density can be
obtained more accurately in one step using MCMC methods. This adaptive importance 
sampling estimator is then used to estimate the marginal likelihood.

To motivate the cross-entropy method, first note that there exists an importance 
sampling density that gives a zero-variance estimator of the marginal likelihood. 
In particular, it is easy to verify that if we use the posterior distribution as 
the importance sampling density, i.e., 
$g^*(\vtheta) = p(\vtheta\gvn\by) = p(\by\gvn\vtheta)p(\vtheta)/p(\by)$,
then the associated importance sampling estimator~\eqref{eq:ISml} has zero variance. 
Of course in practice $g^*$ cannot be used as its normalizing constant is exactly 
the marginal likelihood, the unknown quantity we wish to estimate. 
Nevertheless, this provides a signpost to obtain an optimal importance sampling density. 
Intuitively, if we choose an importance sampling density $g$ that is `close enough' to 
$g^*$ so that both behave similarly, the resulting importance
sampling estimator should have reasonable accuracy. Hence, our goal is to locate a convenient
density that is in a well-defined sense `close' to $g^*$.

To that end, consider a parametric family $\mathcal{F} = \{ f(\vect{\theta};\bv) \}$
indexed by the parameter vector $\bv$ within which we locate the optimal importance sampling density.
We find the density $f(\vtheta;\bv^*)\in\mathcal{F}$ such that it is the `closest' to
$g^*$. One convenient measure of closeness between densities
is the \emph{Kullback-Leibler divergence} or the \emph{cross-entropy distance}.
Specifically, let $f_1$ and $f_2$ be two probability density functions.
Then, the cross-entropy distance from $f_1$ to $f_2$ is defined as:
\begin{equation*}
	\mathcal{D}(f_1,f_2) = \int f_1(\bx)\log \frac{f_1(\bx)}{f_2(\bx)}\di\bx.
\end{equation*}
Given this measure, we can then locate the density $f(\cdot;\bv)\in\mathcal{F}$
such that $\mathcal{D}(g^*,f(\cdot;\bv))$ is minimized: 
$\bv_{\text{ce}}^* = \argmin_{\bv}\mathcal{D}(g^*,f(\cdot;\bv))$. It can be shown that
solving the CE minimization problem is equivalent to finding
\[
	\bv^*_{\text{ce}} = \argmax_{\mathbf{v}}\int p(\by\gvn\vtheta)p(\vtheta)\log f(\vtheta;\bv)\di\vtheta.
\]

This optimization problem is often difficult to solve analytically.
Instead, we consider its stochastic counterpart:
\begin{equation} \label{eq:maxMC}
	\widehat{\bv}^*_{\text{ce}} = \argmax_{\bv} \frac{1}{M}\sum_{m=1}^M \log f(\vect{\theta}_m; \bv),
\end{equation}
where $\vect{\theta}^{(1)},\ldots, \vect{\theta}^{(M)}$ are  draws from the posterior distribution
$p(\vtheta\gvn\by) \propto p(\by\gvn\vtheta)p(\vtheta)$.  

In other words, $\widehat{\bv}^*_{\text{ce}}$ is exactly the maximum likelihood
estimate for $\bv$ if we treat $f(\vect{\theta};\bv)$ as the likelihood function
with parameter vector $\bv$ and $\vect{\theta}^{(1)},\ldots, \vect{\theta}^{(M)}$
an observed sample. Since finding the maximum likelihood estimator is a standard problem,
solving~\eqref{eq:maxMC} is typically easy. In particular, analytical solutions
to \eqref{eq:maxMC} can be found explicitly for the exponential family \citep[e.g.,][p. 70]{rk:ce}.

The parametric family $\mathcal{F}$ is often chosen so that each member 
$f(\vtheta ; \bv)$ is a  product of densities, e.g.,
$f(\vtheta ; \bv) = f(\vtheta_1; \bv_1)\times\cdots\times f(\vtheta_B; \bv_B)$,
where $\vtheta=(\vtheta_1,\ldots, \vtheta_B)$ and $\bv=(\bv_1,\ldots,\bv_B)$.
In that case, one can reduce the possibly high-dimensional maximization problem~\eqref{eq:maxMC}
into $B$ low-dimensional problems, which can then be readily solved.
Once the optimal density is obtained, we then set $g(\cdot) = f(\cdot;\widehat{\bv}^*_{\text{ce}})$ and use
it to construct the importance sampling estimator in \eqref{eq:ISml}.

\subsection{Conditional Monte Carlo}\label{ss:CMC}

Recall that the marginal likelihood can be written as $p(\by) = \Em p(\by\gvn\vtheta)$, 
where $\by$ is fixed and the expectation is taken with respect to the prior distribution $p(\vtheta)$. Suppose that there is a random vector $\vpsi$ with density function  $p(\vpsi)$ such that the conditional expectation $\Em[p(\by\gvn\vtheta)\gvn \vpsi]$ can be computed analytically. By the law of iterated expectation, we have
\[
	p(\by) = \Em[\Em[p(\by\gvn\vtheta)\gvn \vpsi ]],
\]
where the outer expectation is taken with respect to $p(\vpsi)$ and the inner expectation is taken with respect to the conditional distribution $p(\vtheta\gvn\vpsi)$. 
Hence, $\Em[p(\by\gvn\vtheta)\gvn \vpsi]$ is an unbiased estimator of $p(\by)$. In addition, by the variance decomposition formula, we obtain
\[
	\Var (p(\by\gvn\vtheta)) = \Em[\Var(p(\by\gvn\vtheta)\gvn\vpsi)] + \Var(\Em[p(\by\gvn\vtheta)\gvn\vpsi]).
\]
Since $\Em[\Var(p(\by\gvn\vtheta)\gvn\vpsi)] > 0$ if the density $p(\vtheta\gvn\vpsi)$ is non-degenerate, we have
\[
	\Var(\Em[p(\by\gvn\vtheta)\gvn\vpsi]) < \Var(p(\by\gvn\vtheta)).
\]
Therefore, the conditional Monte Carlo estimator $\Em[p(\by\gvn\vtheta)\gvn \vpsi]$ \emph{always} provides variance reduction. The degree of variance reduction, of course, depends on the dimension and variability of $\vpsi$. Loosely speaking, to maximize variance reduction, one feasible approach is to find the `smallest' subset $\vpsi$ of $\vtheta$ such that $\Em[p(\by\gvn\vtheta)\gvn\vpsi]$ is computable. 

In our VARs the parameter vector $\vtheta$ contains the VAR coefficients and other latent variables such as the stochastic volatility and dynamic factors. To have the `smallest' $\vpsi$ is equivalent to integrating out as many elements in $\vtheta$ as possible. In particular, since there are a large number of VAR coefficients in our setting, ideally we would like to remove their contribution to the variability in the marginal likelihood estimation. Fortunately, for all the VARs we consider, we are able to integrate them out analytically and obtain an analytical expression of $\Em[p(\by\gvn\vtheta)\gvn \vpsi]$. We will give an explicit example in the following section.

\subsection{An Example: Marginal Likelihood of VAR-CSV}\label{ss:CSV}

As an illustration, we now provide the details of estimating the marginal likelihood of the VAR-CSV using the proposed method by combining adaptive importance sampling and conditional Monte Carlo. Other VARs with stochastic volatility can be handled similarly and the technical details are given in Appendix A.

Let $\bx_t' = (1, \by_{t-1}',\ldots, \by_{t-p}')$ be a $1\times k$ vector of an intercept and lags with $k=1+np$. Then, stacking the observations over $t=1,\ldots, T$, we can rewrite 
\eqref{eq:VAR}--\eqref{eq:csv} as:
\[
	\bY = \bX\bA + \vepsilon, \quad 
	\text{vec}(\vepsilon)\sim\distn{N}(\mathbf{0},\vSigma\otimes \bD_{\bh}),
\]
where $\bD_{\bh} = \text{diag}(\e^{h_1},\ldots, \e^{h_T})$, 
$\bA = (\ba_0, \bA_1, \ldots, \bA_p)'$ is $k\times n$, and the matrices $\bY$, $\bX$ and 
$\vepsilon$ are respectively of dimensions $T\times n$, $T\times k$ and $T\times n$. In addition, the natural conjugate prior on $(\bA, \vSigma)$ depends on a single hyperparameter 
$\kappa$ that controls the overall shrinkage strength, which is treated as an unknown parameter. Finally, the log-volatility $h_t$ follows the AR(1) process in \eqref{eq:h} with AR coefficient $\phi$ and error variance $\sigma^2$. 

Next, we define an appropriate conditional Monte Carlo estimator, which will then be combined with adaptive importance sampling. Using the notations in Section~\ref{ss:CMC}, the set of model parameters and latent variables is $\vtheta = \{\bA, \vSigma, \phi, \sigma^2, \kappa, \bh\}$, and the conditional likelihood (i.e., conditional on the latent variables $\bh$) is given by $p(\by\gvn \vtheta) = p(\bY \gvn \bh, \bA, \vSigma, \phi, \sigma^2, \kappa) = p(\bY \gvn \bh, \bA, \vSigma)$.  When $n$ is large, the dimensions of both $\bA$ and $\vSigma$ are large, and we expect them to contribute the most to the variability of $p(\bY \gvn \bh, \bA, \vSigma)$. Hence, we aim to integrate them out. To that end, we let $\vpsi = \{\bh, \phi, \kappa \}$ and consider the conditional expectation $\Em[p(\by\gvn \vtheta)\gvn \vpsi] = p(\bY\gvn\bh,\kappa)$, which can be computed analytically:
\begin{align}
	p(\bY\gvn\bh,\kappa) & = \int p(\bY\gvn\bh,\bA,\vSigma)p(\bA,\vSigma\gvn\kappa) \di(\bA,\vSigma)\nonumber\\
	& = \pi^{-\frac{Tn}{2}}\e^{-\frac{n}{2}\mathbf{1}_T'\bh}
	|\bV_{\bA}|^{-\frac{n}{2}}|\bK_{\bA}|^{-\frac{n}{2}}
	\frac{\Gamma_n\left(\frac{\nu_0+T}{2}\right)|\bS_0|^{\frac{\nu_0}{2}}}
	{\Gamma_n\left(\frac{\nu_0}{2}\right)|\hat{\bS}|^{\frac{\nu_0+T}{2}}}, \label{eq:ygvnh}
\end{align}
where $\Gamma_n(\cdot)$ is the multivariate gamma function and 
\begin{align*}
	\bK_{\bA} & = \bV_{\bA}^{-1} + \bX'\bD^{-1}_{\bh} \bX, \quad
		\hat{\bA}  = \bK_{\bA}^{-1}(\bV_{\bA}^{-1}\bA_0 +  \bX'\bD^{-1}_{\bh} \bY), \\
	\hat{\bS} & = \bS_0 + \bA_0'\bV_{\bA}^{-1}\bA_0 + \bY'\bD^{-1}_{\bh}\bY
	-\hat{\bA}'\bK_{\bA}\hat{\bA}.
\end{align*}
Note that the overall shrinkage parameter $\kappa$ appears in the prior covariance matrix $\bV_{\bA}$. The details of the derivations are given in Appendix A. We note that~\eqref{eq:ygvnh} can be calculated quickly without inverting any large matrices; see, e.g., \citet{chan20} for various computational shortcuts. 

Now, the marginal likelihood can be written in terms of $p(\bY\gvn\bh,\kappa)$ as follows:
\[
	p(\by) = \int p(\bY\gvn\bh,\kappa)p(\bh\gvn\phi,\sigma^2)p(\phi,\sigma^2,\kappa)
	\di(\bh,\phi,\sigma^2,\kappa),
\]
where $p(\bh\gvn\phi,\sigma^2)$ is a Gaussian density implied by the state equation \eqref{eq:h}
and $p(\phi,\sigma^2,\kappa)$ is the prior for $(\phi,\sigma^2,\kappa)$. Hence, the corresponding conditional Monte Carlo estimator is given by
\[
	\hat{p(\by)}_{\rm CMC} = \frac{1}{M}\sum_{m=1}^M p(\bY\gvn\bh^{(m)}, \kappa^{(m)}),
\]
where $\bh^{(1)},\ldots, \bh^{(M)}$ are drawn from the prior $p(\bh) = \int p(\bh\gvn\phi,\sigma^2)p(\phi,\sigma^2)\di(\phi,\sigma^2)$ and $\kappa^{(1)},\ldots, \kappa^{(M)}$ are from the prior $p(\kappa)$. 

The conditional Monte Carlo estimator $\hat{p(\by)}_{\rm CMC}$ alone does not work well in practice, because most draws from the prior distributions $p(\bh)$ and $p(\kappa)$ do not coincide with the high-density region of $p(\bY\gvn\bh,\kappa)$ (as a function of $\bh$ and $\kappa$). To tackle this problem, next we combine the conditional Monte Carlo approach with importance sampling, where the importance sampling density is obtained using the improved cross-entropy method. \citet{CE15} first propose using the improved cross-entropy method to estimate the marginal likelihood, but the models considered there are low dimensional. In contrast, for our settings with a large number of latent variables, additional consideration on parsimonious parameterization is required for obtaining appropriate importance sampling densities. 

More specifically, we consider the parametric family
\[
	\mathcal{F} = \{f_{\distn{N}}(\bh; \hat{\bh}, \hat{\bK}_{\bh}^{-1})
	f_{\distn{N}}(\phi; \hat{\phi}, \hat{K}_{\phi}^{-1})
	f_{\distn{G}}(\kappa; \hat{\nu}_{\kappa}, \hat{S}_{\kappa})	
	\},
\]
where $f_{\distn{N}}$ and $f_{\distn{G}}$ denote Gaussian and gamma densities, respectively.
Moreover, $\hat{\bh}$ and $\hat{\bK}_{\bh}$ are respectively the mean vector and precision matrix---i.e., inverse covariance matrix---of the $T$-variate Gaussian density of $\bh$;  $\hat{\phi}$ and 
$\hat{K}_{\phi}$ are the mean and precision of the univariate Gaussian density of $\phi$; and 
$\hat{\nu}_{\kappa}$ and $\hat{S}_{\kappa}$ are respectively the shape and rate of the gamma importance density for $\kappa$.\footnote{Note that $\sigma^2$ only appears in the prior of $\bh$, and one can integrate it out analytically: $p(\bh) = \int p(\bh\gvn\phi,\sigma^2)p(\phi,\sigma^2)\di(\phi,\sigma^2) = \int p(\bh\gvn\phi)p(\phi)\di\phi$, where $p(\bh\gvn\phi)$ has an analytical expression, which is given in Appendix~A. Hence, there is no need to simulate $\sigma^2$.} Now, we aim to choose these parameters so that the associated member in $\mathcal{F}$ is the closest in cross-entropy distance to the theoretical zero-variance importance sampling density $p(\bh,\phi,\kappa\gvn \by)\propto p(\bY\gvn\bh,\kappa)p(\bh\gvn\phi)p(\phi, \kappa)$, which is simply the marginal posterior density. Draws from this marginal distribution can be obtained using the posterior sampler described in Appendix~A---i.e., we obtain posterior draws from the full posterior 
$p(\bA,\vSigma,\bh,\phi,\sigma^2,\kappa \gvn \by)$ and keep only the draws of 
$\bh,\phi,$ and~$\kappa$.
 
Now, given the posterior draws $(\bh^{(1)},\phi^{(1)},\kappa^{(1)}) ,\ldots, (\bh^{(M)},\phi^{(M)},\kappa^{(M)})$ and the parametric family $\mathcal{F}$, the optimization problem in \eqref{eq:maxMC} can be divided into 3 lower-dimensional problems: 1) obtain $\hat{\bh}$ and $\hat{\bK}_{\bh}$ given the posterior draws of $\bh$; 2) obtain $\hat{\phi}$ and $\hat{K}_{\phi}$ given the posterior draws of $\phi$; and 3) obtain $\hat{\nu}_{\kappa}$ and $\hat{S}_{\kappa}$ given the posterior draws of $\kappa$. The latter two steps are easy as both densities are univariate, and they can be solved using similar methods as described in \citet{CE15}. Below we focus on the first step.

If we do not impose any restrictions on the Gaussian density $f_{\distn{N}}(\bh; \hat{\bh}, \hat{\bK}_{\bh}^{-1})$, in principle we can obtain $\hat{\bh}$ and $\hat{\bK}_{\bh}$ analytically given the posterior draws $\bh^{(1)} ,\ldots, \bh^{(M)}$. However, there are two related issues with this approach, First, if unrestricted, the precision matrix $\hat{\bK}_{\bh}$ is a full, $T\times T$ matrix. Evaluating and sampling from  this Gaussian density would be time-consuming. Second, since $\hat{\bK}_{\bh}$ is a symmetric but otherwise unrestricted matrix, there are $T(T+1)/2$ parameters to be estimated. Consequently, one would require a large number of posterior draws to ensure that $\hat{\bh}$ and $\hat{\bK}_{\bh}$ are accurately estimated.

In view of these potential difficulties, we consider instead a restricted family of Gaussian densities that exploits the time-series structure. Specifically, this family is parameterized by the parameters $\rho$, $\ba = (a_1,\ldots, a_T)'$ and $\mathbf{b} = (b_1,\ldots, b_T)'$ as follows. First, $h_1$ has the marginal Gaussian distribution $h_1\sim\distn{N}(a_1,b_1)$. For $t=2,\ldots, T,$ 
\begin{equation} \label{eq:h_IS}
	h_t = a_t + \rho h_{t-1} + \eta_t, \quad \eta_t\sim\distn{N}(0,b_t).
\end{equation}
In other words, the joint distribution of $\bh$ is implied by an AR(1) process with time-varying intercepts and variances. It is easy to see that this parametric family includes 
the prior density of $\bh$ implied by the state equation~\eqref{eq:h} as a member.\footnote{We also investigate other Gaussian families such as the joint distributions implied by the AR(2), MA(1) and ARMA(1,1,) processes. None of them perform noticeably better than the simple AR(1) process in \eqref{eq:h_IS}.}

To facilitate computations, we vectorize the AR(1) process and write
\[
	\bH_{\rho} \bh = \ba + \veta, \quad \veta\sim\distn{N}(\mathbf{0}, \mathbf{B}),
\]
where $\mathbf{B}=\text{diag}(b_1,\ldots, b_T)$ and 
\[
	\bH_{\rho} =
	\begin{pmatrix}
	1 & 0 & \cdots & 0 \\
	-\rho & 1 & \cdots & 0 \\ 	
	\vdots & \ddots  & \ddots & \vdots \\
	0 & \cdots & -\rho & 1
	\end{pmatrix}.
\]
Since $|\bH_{\rho}|=1$ for any $\rho$, $\bH_{\rho}$ is invertible. Then, the Gaussian distribution implied by the AR(1) process in \eqref{eq:h_IS} has the form $\bh\sim\distn{N}(\bH_{\rho}^{-1} \ba, (\bH_{\rho}'\mathbf{B}^{-1}\bH_{\rho})^{-1})$.
Note that the number of parameters here is only 2$T$ + 1 instead of unrestricted case of $T + T(T+1)/2$.

Next, given this parametric family and posterior draws $\bh^{(1)} ,\ldots, \bh^{(M)}$, we can solve the maximization problem~\eqref{eq:maxMC} in two steps. First, note that given $\rho$, we can obtain the maximizer 
$\hat{\ba} = (\hat{a}_1,\ldots, \hat{a}_T)'$ and  $\hat{\mathbf{b}} = (\hat{b}_1,\ldots, \hat{b}_T)'$ analytically. More specifically, by maximizing the log-likelihood 
\[
	\ell(\rho,\ba, \mathbf{b}) = -\frac{TM}{2}\log(2\pi) - \frac{M}{2}\log|\mathbf{B}| 
	-\frac{1}{2}\sum_{m=1}^M(\bH_{\rho}\bh^{(m)}-\ba)'\mathbf{B}^{-1}
	(\bH_{\rho}\bh^{(m)}-\ba)
\]
with respective to $\ba$ and $\mathbf{b}$, we obtain the maximizer 
 $\hat{\ba} = \frac{1}{M}\sum_{m=1}^M\bH_{\rho}\bh^{(m)}$	and 
$\hat{b}_t = \frac{1}{M}\sum_{m=1}^M(h^{(m)}_t - \hat{a}_t)^2, t=1,\ldots, T$. 

Second, given the analytical solution $\hat{\ba}$ and $\hat{\mathbf{b}}$---which are functions of $\rho$---we can readily evaluate the one-dimensional concentrated log-likelihood $\ell(\rho,\hat{\ba}, \hat{\mathbf{b}})$. Then, $\hat{\rho}$ can be obtained numerically by maximizing $\ell(\rho,\hat{\ba}, \hat{\mathbf{b}})$ with respect to $\rho$. Finally, we use $f_{\distn{N}}(\bh; \bH_{\hat{\rho}}^{-1}\hat{\ba}, (\bH_{\hat{\rho}}'\hat{\mathbf{B}}^{-1}\bH_{\hat{\rho}})^{-1})$, where $\hat{\mathbf{B}} = \text{diag}(\hat{b}_1,\ldots, \hat{b}_T)$, as the importance sampling density for $\bh$.

\citet{CE18} also consider a Gaussian importance sampling density for the log-volatility
$\bh$. However, there the approximation is based on a second-order Taylor expansion at the mode. Hence, it is a local approximation that might not be close to the target density at points away from the mode. In contrast, our proposed Gaussian importance sampling density is a global approximation that takes into account of the whole support of the distribution, and is therefore expected to behave more similarly to the target distribution $p(\bh\gvn \by)$ over its entire support.

\section{Monte Carlo Experiments} \label{s:MC}

In this section we conduct a series of Monte Carlo experiments to assess how the proposed marginal likelihood estimators perform in selecting the correct data generating process. More specifically, we focus on the following three questions. First, can we distinguish the three stochastic volatility models described in Section~\ref{s:models}? Second, can we discriminate between models with time-varying volatility against homoskedastic errors?  Finally, for factor stochastic volatility models, can we identify the correct number of factors?
 
\subsection{Can We Distinguish the Stochastic Volatility Specifications?}

In the first part of this Monte Carlo exercise, we address the question of whether one can distinguish the three stochastic volatility models---namely, the common stochastic volatility model (VAR-CSV), the Cholesky stochastic volatility model (VAR-SV) and the factor stochastic volatility model (VAR-FSV)---in the context of large Bayesian VARs. To that end, we generate 100 datasets from each of the three models. Each dataset consists of $n=10$ variables, $T = 400$ observations and $p=2$ lags. Given each dataset, we compute the log marginal likelihoods of the same three models using the proposed method described in Section~\ref{s:ML}.

\begin{figure}[H]
  \centering
  \includegraphics[width=.8\textwidth]{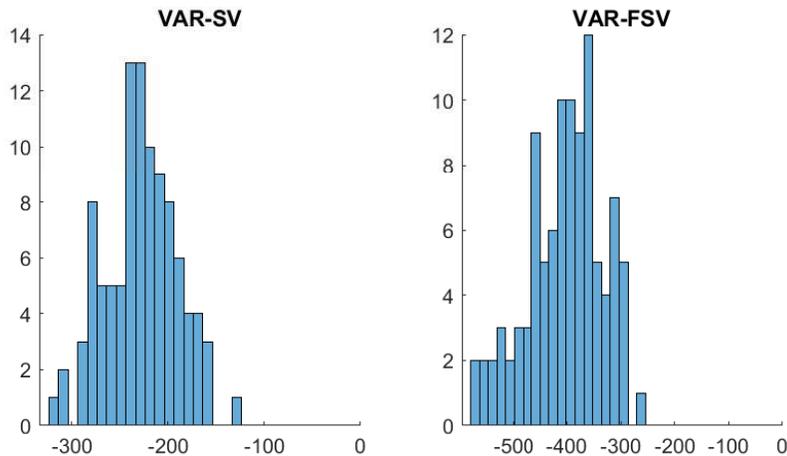}
  \caption{Histograms of log marginal likelihoods under VAR-SV (left panel) and VAR-FSV (right panel) relative to the true model (VAR-CSV). A negative value indicates that the correct model is favored.}
   \label{fig:MC1-CSV}
\end{figure}

For VAR-CSV, we generate the intercepts from $\distn{U}(-10,10)$. The diagonal elements of the first VAR coefficient matrix are iid $\distn{U}(-0.2, 0.4)$ and the off-diagonal elements are $\distn{U}(-0.2, 0.2)$; all elements of the second VAR coefficient matrix are iid $\distn{N}(0,0.05^2)$. The error covariance matrix $\vSigma$ is generated from the inverse-Wistart distribution $\distn{IW}(n+5,0.7\times \mathbf{I}_n + 0.3\times\mathbf{1}_n\mathbf{1}_n')$, where $\mathbf{1}_n$ is an $n\times 1$ column of ones. For the log-volatility process, we set $\phi = 0.98$ and $\sigma^2 = 0.1$. For VAR-SV, the VAR coefficients are generated as before, and the free elements of the impact matrix are generated independently from the $\distn{N}(0,0.5^2)$ distribution. For the log-volatility process, we set $\mu_i=-1$, $\phi_i = 0.98$ and $\sigma^2_i = 0.1$ for $i=1,\ldots, n$. Finally, for VAR-FSV, we set the number of factor $r$ to be 3. The VAR coefficients are generated the same way as other models.  For the log-volatility process, we set $\mu_i=-1$, $\phi_i = 0.98$ and $\sigma^2_i = 0.1$ for $i=1,\ldots, n$, and $\mu_{n+j} = 0$, $\phi_{n+j} = 0.98$ and $\sigma^2_{n+j} = 0.1$ for $j=1,\ldots, r$. That is, the log stochastic volatility processes associated with the factors have a larger mean, but are otherwise the same as the idiosyncratic stochastic volatility processes.\footnote{Since the main goal of the Monte Carlo experiments is to assess if one can distinguish the stochastic volatility specifications, we aim to use similar priors across the models so as to minimize their impacts. In particular, for the hierarchical Minnesota priors we fix all shrinkage hyperparameters to be $\kappa =  \kappa_1 = \kappa_2 = 0.2^2$. That is, we turn off the cross-variable shrinkage feature of the Minnesota priors under VAR-SV and VAR-FSV, so that they are comparable to the symmetric Minnesota prior under VAR-CSV.}
In the first experiment, we generate 100 datasets from VAR-CSV as described above. For each dataset, we then compute the log marginal likelihoods of VAR-SV and VAR-FSV relative to that of the true model VAR-CSV. Specifically, we subtract the latter log marginal likelihood from the log marginal likelihoods of VAR-SV and VAR-FSV. The results are reported in Figure~\ref{fig:MC1-CSV}. 

Since a model is preferred by the data if it has a larger log marginal likelihood value, a difference that is negative indicates that the correct model is favored. It is clear from the histograms that for all the datasets the correct model VAR-CSV compares favorably to the other two stochastic volatility specifications, often by a large margin.

\begin{figure}[H]
  \centering
  \includegraphics[width=.8\textwidth]{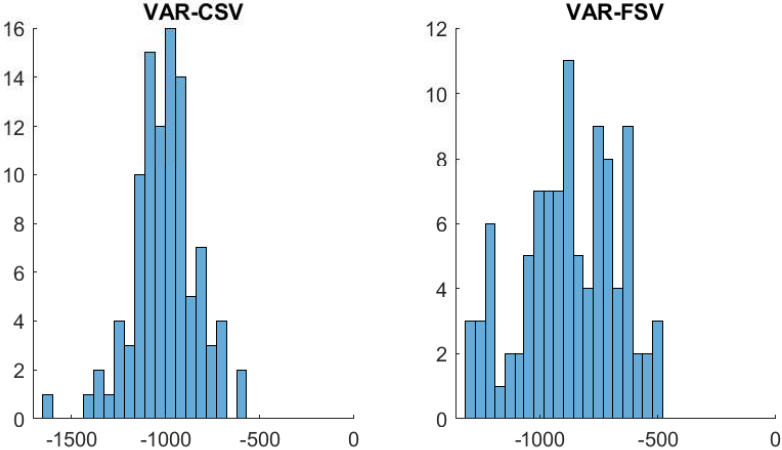}
  \caption{Histograms of log marginal likelihoods under VAR-CSV (left panel) and VAR-FSV (right panel) relative to the true model (VAR-SV). A negative value indicates that the correct model is favored.}
   \label{fig:MC1-SV}
\end{figure}

Next, we generate 100 datasets from VAR-SV. For each dataset, we then compute the log marginal likelihoods of VAR-CSV and VAR-FSV relative to that of the true model. The results are reported in Figure~\ref{fig:MC1-SV}. Again, the model comparison result shows that, for all datasets, the correct model VAR-SV is overwhelmingly favored compared to the other two stochastic volatility specifications. 

\begin{figure}[H]
  \centering
  \includegraphics[width=.8\textwidth]{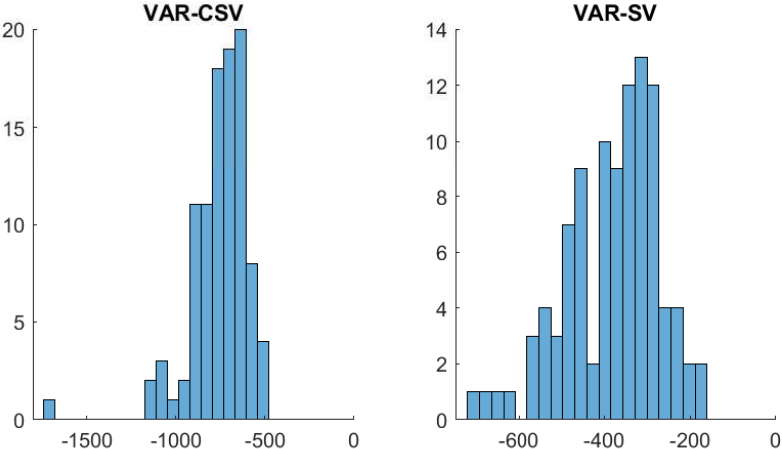}
  \caption{Histograms of log marginal likelihoods under VAR-CSV (left panel) and VAR-SV (right panel) relative to the true model (VAR-FSV). A negative value indicates that the correct model is favored.}
   \label{fig:MC1-FSV}
\end{figure}

Lastly, we generate 100 datasets from VAR-FSV, and for each dataset we compute the log marginal likelihoods of VAR-CSV and VAR-SV relative to that of VAR-FSV. The results are reported in Figure~\ref{fig:MC1-FSV}. Once again, these results show that the proposed marginal likelihood estimators can select the correct model for all the simulated datasets.

Appendix C provides additional Monte Carlo results to further assess the performance of the proposed marginal likelihood estimators in higher dimensions ($n=20$) and noisier VAR equations (magnitudes of the error covariance matrices increase 10-fold). For these additional settings, the correct model is again overwhelmingly favored compared to the other two stochastic volatility models. All in all, this series of Monte Carlo experiments show that using the proposed marginal likelihood estimators one can clearly distinguish the three stochastic volatility specifications, even for a moderate number of variables and sample size. 

\subsection{Can We Discriminate between Models with Time-Varying Volatility against Homoskedastic Errors?}

\begin{figure}[H]
  \centering
  \includegraphics[width=1\textwidth]{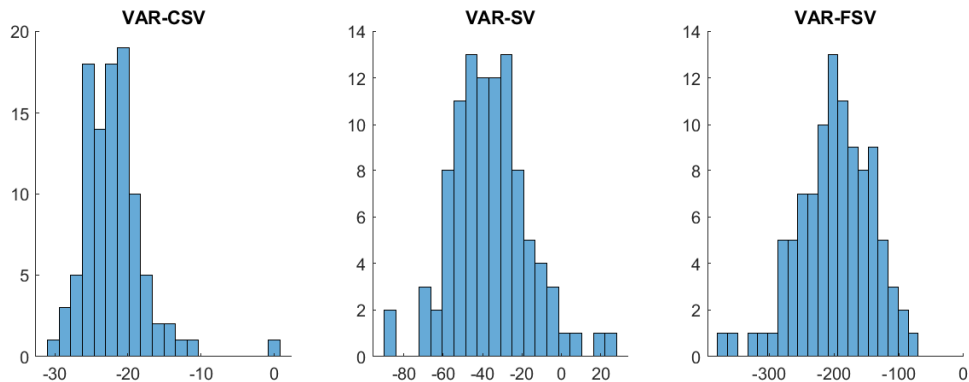}
  \caption{Histograms of log marginal likelihoods under VAR-CSV (left panel), VAR-SV (middle panel) and VAR-FSV (right panel) relative to the true model (homoskedastic VAR). A negative value indicates that the correct model is favored.}
   \label{fig:MC2-VAR}
\end{figure}

Since VARs with stochastic volatility are by design more flexible than conventional VARs with homoskedastic errors, one concern of using these more flexible models is that they might overfit the data. In this Monte Carlo experiment we investigate if the proposed marginal likelihood estimators can distinguish models with and without stochastic volatility. More specifically, we generate 100 datasets from a standard VAR with homoskedasic errors. For each dataset, we then compute the log marginal likelihoods of the three stochastic volatility models: VAR-CSV, VAR-SV and VAR-FSV.\footnote{We also generate data from the stochastic volatility models and compare their marginal likelihoods with a homoskedastic VAR. The results overwhelmingly favor the stochastic volatility models as the homoskedastic VAR does not fit the data well at all. For space constraint, however, we do not report these model comparison results.} The differences in log marginal likelihoods relative to the homoskedastic VAR are reported in Figure~\ref{fig:MC2-VAR}.

Recall that a model is preferred by the data if it has a larger log marginal likelihood value. Hence, a negative value indicates that the correct homoskedastic VAR is selected. Overall, the histograms show that for almost all datasets the correct homoskedastic model compares favorably to the three stochastic volatility models. This Monte Carlo experiment highlights that the marginal likelihood does not always favor the model with the best model-fit; it also has a built-in penalty for model complexity. And it is only when the gain in model-fit outweighs the additional model complexity does the more complex model have a larger marginal likelihood value.

It is also interesting to note that VAR-CSV generally performs better than the other two stochastic volatility models. This is perhaps not surprising, as the VAR-CSV is the most parsimonious extension of the homoskedastic VAR---it has only one stochastic volatility process $h_t$, and when $h_t$ is identically zero it replicates the homoskedastic VAR. Nevertheless, even for this closely related extension, the marginal likelihood criterion clearly indicates that the stochastic volatility process is spurious and it tends to favor the homoskedastic model.

\subsection{Can We Identify the Correct Number of Factors?}

One important specification choice for factor stochastic volatility models is to select the number of factors. This represents the trade-off between a more parsimonious model with fewer stochastic volatility factors versus a more complex model with more factors but better model-fit. In this Monte Carlo experiment, we investigate if the proposed marginal likelihood estimator for the VAR-FSV can pick the correct number of factors. 

\begin{figure}[H]
  \centering
  \includegraphics[width=.8\textwidth]{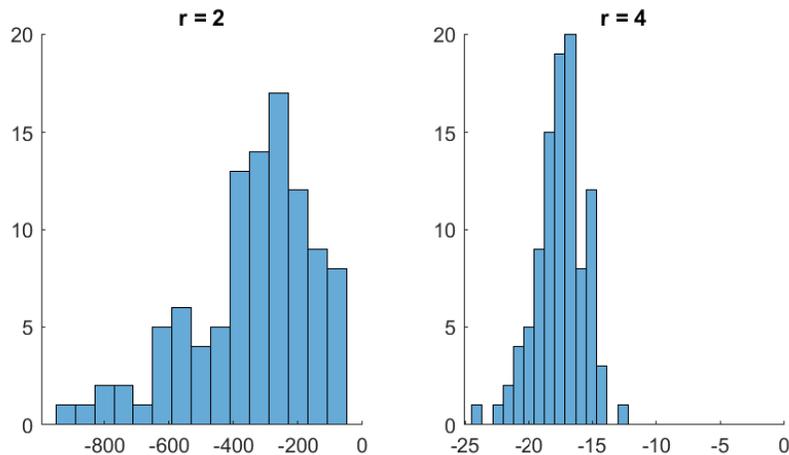}
  \caption{Histograms of log marginal likelihoods under the 2-factor model (left panel) and the 4-factor model (right panel) relative to the true 3-factor model. A negative value indicates that the correct model is favored.}
   \label{fig:MC3-FSV}
\end{figure}

To that end, we generate 100 datasets from VAR-FSV with $r=3$ factors. Then, for each dataset, we compute the log marginal likelihood of VAR-FSV models with $r=2,3$ and 4 factors. This set of number of factors is chosen to shed light on the effects of under-fitting versus over-fitting. We report the log marginal likelihoods of the 2- and 4-factor models relative to the true 3-factor model in Figure~\ref{fig:MC3-FSV}. The results clearly show that the proposed method is able to identify the correct number of factors. In particular, for all datasets the 3-factor model outperforms the more parsimonious 2-factor model and the more flexible 4-factor model.

The 2-factor model under-fits the data and performs much worse than the correct 3-factor model. On the other hand, while the 4-factor model is able to replicate features of the 3-factor model, it also includes spurious features that tend to over-fit the data. Consequently, the 3-factor model is strongly preferred by the data over the two alternatives. However, the impacts of under- and over-fitting are not symmetric. In particular, under-fitting receives a much heavier penalty than over-fitting, as illustrated by the much lager differences (in magnitude) between the 2- and 3-factor models than those between the 3- and 4-factor models. Regardless of these differences, we conclude that the proposed method is able to select the correct number of factors.

\section{Empirical Application}\label{s:application}

In this empirical application we compare the three stochastic volatility specifications---i.e., the common stochastic volatility model (VAR-CSV), the Cholesky stochastic volatility model (VAR-SV) and the factor stochastic volatility model (VAR-FSV)---in the context of Bayesian VARs of different dimensions. We use a dataset that consists of 30 US quarterly variables with a sample period from 1959Q1 to 2019Q4. It is constructed from the FRED-QD database at the Federal Reserve Bank of St. Louis as described in \citet{MN20}. The dataset contains a range of widely used macroeconomic and financial variables, including Real GDP and its components, various measures of inflation, labor market variables and interest rates. They are transformed to stationarity, typically to annualized growth rates. We consider VARs that are small ($n=7$), medium ($n=15$) and large ($n=30$). The complete list of variables for each dimension and how they are transformed is given in Appendix~B. 

In contrast to the Monte Carlo study where the shrinkage hyperparameters in the hierarchical Minnesota priors are fixed at certain subjective values, here we treat them as unknown parameters to be estimated. This is motivated by a few recent papers that find this data-based approach outperforms subjective prior elicitation in both in-sample model fit and out-of-sample forecast performance \citep[see, e.g.,][]{CCM15,GLP15,AMW20}. For easy comparison, we set the lag length to be $p=4$ for all VARs.\footnote{The lag length can of course be chosen by comparing the marginal likelihood. In a preliminary analysis, we find that a lag length of $p=4$ is generally sufficient. In addition, with our flexible hierarchical Minnesota priors on the VAR coefficients, adding more lags than necessary does not substantially affect model performance. For instance, for all of the three stochastic volatility specifications, the best models for $n=7$ have $p=3$ lags. Adding one more lag reduces the log marginal likelihood by only about 1-2 for all models.} We compute the log marginal likelihoods of the VARs using the proposed hybrid algorithm described in Section \ref{s:ML}. The importance sampling density for each model is constructed using the cross-entropy method with 20,000 posterior draws after a burn-in period of 1,000. Then, the log marginal likelihood estimate is computed using an importance sample of size 10,000.

\subsection{Comparing Stochastic Volatility Specifications}

We first report the main model comparison results on comparing the three stochastic volatility specifications: VAR-CSV, VAR-SV and VAR-FSV. As a benchmark, we also include the standard homoskedastic VAR with the natural conjugate prior (its marginal likelihood is available analytically). In addition to the widely different likelihoods implied by these stochastic volatility specifications, they also differ in the flexibility of the shrinkage priors employed. More specifically, both the standard VAR and VAR-CSV use the natural conjugate prior, which has only one hyperparameter that controls the shrinkage strength of all VAR coefficients. In contrast, both the priors under VAR-SV and VAR-FSV can accommodate cross-variable shrinkage. That is, there are two separate shrinkage hyperparameters, one controls the shrinkage strength on own lags, whereas the other controls that of lags of other variables. In what follows we focus on the overall ranking of the models; we will investigate the role of the priors in the next section.

Table \ref{tab:ML1} reports the log marginal likelihood estimates of the four VARs across the three model dimensions ($n=7, 15, 30$). First, it is immediately apparent that all three stochastic volatility models perform substantially better than the standard homoskedastic VAR. For example, the log marginal likelihood difference between VAR-CSV and VAR is about 760 for $n=30$, highlighting overwhelming evidence in favor of the heteroskedastic model. This finding is in line with the growing body of evidence that shows the importance of time-varying volatility in modeling both small and large macroeconomic datasets. 

\begin{table}[H]
\caption{Log marginal likelihood estimates (numerical standard errors) of a standard homoskedastic VAR, VAR-CSV, VAR-SV and VAR-FSV.}
\label{tab:ML1}
\centering
\begin{tabular}{lcccccc}
\hline\hline
	&	VAR	&	VAR-CSV	&	VAR-SV	&	\multicolumn{3}{c}{VAR-FSV}	\\	 \hline
	&		&		&		&	$k=1$	&	$k=2$	&	$k=3$	\\ 
\rowcolor{lightgray}
$n=7$	&	$-$2,583.4	&	$-$2,410	&	$-$2,315	&	$-$2,318	&	$-$2,320	&	$-$2,331	\\
\rowcolor{lightgray}
	&	-	&	(0.1)	&	(0.3)	&	(0.3)	&	(0.7)	&	(0.7)	\\
	&		&		&		&	$k=3$	&	$k=4$	&	$k=5$	\\
	\rowcolor{lightgray}
$n=15$	&	$-$6,918.8	&	$-$6,618	&	$-$6,443	&	$-$6,468	&	$-$6,454	&	$-$6,485	\\
\rowcolor{lightgray}
	&	-	&	(0.1)	&	(0.4)	&	(0.7)	&	(0.8)	&	(1.3)	\\
	&		&		&		&	$k=9$	&	$k=10$	&	$k=11$	\\
	\rowcolor{lightgray}
$n=30$	&	$-$12,783.4	&	$-$12,024	&	$-$11,382 &	$-$11,571	&	$-$11,567	&	$-$11,608	\\
\rowcolor{lightgray}
	&	-	&	(0.1)	&	(0.6)	&	(1.2)	&	(1.8)	&	(1.8)	\\  \hline\hline
\end{tabular}
\end{table}

Second, among the three stochastic volatility specifications, the data overwhelmingly prefers VAR-SV and VAR-FSV over the more restrictive VAR-CSV for all model dimensions, due to a combination of the more flexible likelihoods and priors. For example, the log marginal likelihood differences between VAR-SV and VAR-CSV are 95 for $n=7$, 175 for $n=15$ and 642 for $n=30$. Third, for all three model dimensions, VAR-SV is the most favored stochastic volatility specification, though VAR-FSV comes in close second. Finally, we note that the optimal number of factors changes across the model dimension. It is perhaps not surprising that more factors are needed to model the more complex error covariance structure as the model dimension increases. For instance, for $n=7$ the 1-factor model performs the best, whereas one needs 10 factors when $n=30$.

To corroborate these model comparison results, we next perform a recursive out-of-sample forecasting exercise using an evaluation period from 1975Q1 to the end of the sample. More specifically, Table~\ref{tab:forecasts} reports the joint density forecasting results (as measured by the sum of log predictive likelihoods) for the four VARs across the three model dimensions (for VAR-FSV, we fix $k=4$). It is clear from the table that one can draw mostly similar conclusions from these forecasting results. In particular, all three stochastic volatility models perform substantially better than the homoskedastic VAR for both 1- and 4-step-ahead forecast horizons and for all model dimensions. This again highlights the importance of allowing for some form of time-varying volatility. In addition, for 1-step-ahead, both VAR-SV and VAR-FSV forecast better than the more restrictive VAR-CSV for all model dimensions. However, when the forecast horizon increases, VAR-CSV tends to perform better for higher model dimensions.

\begin{table}[H]
\caption{Sum of log predictive likelihoods of a standard homoskedastic VAR, VAR-CSV, VAR-SV and VAR-FSV. A larger value indicates better forecasting performance.}
\label{tab:forecasts}
\centering
\resizebox{\textwidth}{!}{\begin{tabular}{lcccccccc}
\hline\hline
	&	\multicolumn{4}{c}{1-step-ahead}					&	\multicolumn{4}{c}{4-step-ahead}							\\
	&	VAR	&	VAR-CSV	&	VAR-SV	&	VAR-FSV	&	VAR	&	VAR-CSV	&	VAR-SV	&	VAR-FSV	\\ \hline
$n=7$	&	$-$1,858	&	$-$1,719	&	$-$1,625	&	$-$1,613	&	$-$2,573	&	$-$2,321	&	$-$2,311	&	$-$2,312	\\
\rowcolor{lightgray}
$n=15$	&	$-$5,070	&	$-$4,764	&	$-$4,612	&	$-$4,604	&	$-$6,243	&	$-$5,631	&	$-$5,821	&	$-$5,772	\\
$n=30$	&	$-$11,271	&	$-$8,579	&	$-$8,035	&	$-$8,424	&	$-$15,715	&	$-$10,030	&	$-$11,739	&	$-$11,348	\\
 \hline\hline
\end{tabular}}
\end{table}

\subsection{Comparing Shrinkage Priors}

In this section we compare different types of Minnesota priors for each of the three stochastic volatility specifications. In particular, we investigate the potential benefits of allowing for cross-variable shrinkage and selecting the overall shrinkage hyperparameters in a data-driven manner. To that end, we consider two useful benchmarks. First, for VAR-SV and VAR-FSV we consider the special case where $\kappa_1 = \kappa_2$---i.e., we turn off the cross-variable shrinkage and require the shrinkage hyperparameters on own and other lags to be the same. We refer to this version as the symmetric prior. The second benchmark is a set of subjectively chosen hyperparameter values that apply cross-variable shrinkage. In particular, we follow \citet*{CCM15} and use the values $\kappa_1 = 0.04$ and $\kappa_2 = 0.0016$. This second benchmark is referred to as the subjective prior. Finally, our baseline prior, where $\kappa_1$ and $\kappa_2$ are estimated from the data and could potentially be different, is referred to as the asymmetric prior. 

To fix ideas, we focus on VARs with $n=15$ variables. Table~\ref{tab:ML2} reports the log marginal likelihood estimates of the three stochastic volatility specifications with different shrinkage priors. First, for both VAR-SV	and VAR-FSV, the asymmetric prior significantly outperforms the symmetric version that requires $\kappa_1 = \kappa_2$. This result suggests that it is beneficial to shrink the coefficients on own lags differently than those on lags of other variables. This makes intuitive sense as one would expect that, on average, a variable's own lags would be more important for its future evolution than lags of other variables. By relaxing the restriction that $\kappa_1 = \kappa_2$, the log marginal likelihood values of VAR-SV and VAR-FSV increase by 177 and 204, respectively. 

\begin{table}[H]
\caption{Log marginal likelihood estimates (numerical standard errors) of the stochastic volatility specifications with different shrinkage priors for $n=15$.}
\label{tab:ML2}
\centering
\begin{tabular}{lccc}
\hline\hline	
	&	VAR-CSV	&	VAR-SV	&	VAR-FSV	($k=4$) \\ \hline
Subjective prior	&	$-$6,702	&	$-$6,474	&	$-$6,491	\\
	&	(0.1)	&	(0.6)	&	(0.9)	\\
\rowcolor{lightgray}
Symmetric prior	&	$-$6,618	&	$-$6,620	&	$-$6,658	\\
\rowcolor{lightgray}
	&	(0.1)	&	(0.4)	&	(1.1)	\\
Asymmetric prior	&	-	&	$-$6,443	&	$-$6,454	\\
	&		&	(0.7)	&	(0.8)	\\
	\hline\hline
\end{tabular}
\end{table}

In addition, there are also substantial benefits of allowing the shrinkage hyperparameters to be estimated instead of fixing them subjectively. For example, the log marginal likelihood value of VAR-CSV with the symmetric prior is 84 larger than that of the subjective prior; the log marginal likelihood value of VAR-SV with the asymmetric prior is 31 larger than that of the subjective prior. These results suggest that while those widely-used subjective hyperparameter values seem to work well for some datasets, they might not be suitable for others that contain different variables and span different sample~periods. 

We reported in last section that the data overwhelmingly preferred the more flexible VAR-SV and VAR-FSV over VAR-CSV. But it was unclear whether it was due to the more flexible likelihoods or priors (recall that VAR-CSV can only accommodate the symmetric prior). The result here suggests that the superior performance of VAR-SV and VAR-FSV can be attributed to the more flexible priors rather than the more flexible stochastic volatility specifications. In particular, both VAR-SV and VAR-FSV with the symmetric prior actually perform slightly worse than VAR-CSV (respectively, $-$6,620, $-$6,658 and $-$6,618 for the three models). Only when the asymmetric prior is used do VAR-SV and VAR-FSV outperform VAR-CSV. This stark result shows that choosing a suitable shrinkage prior in high-dimensional settings is as important as, if not more important than, selecting a suitable stochastic volatility specification.

Next, Table~\ref{tab:kappas} reports the posterior estimates of the shrinkage hyperparameters under the three stochastic volatility specifications. First, under VAR-CSV, the posterior mean of $\kappa$ varies across the model dimensions, from 0.32 for $n=7$ to 0.22 for $n=15$ and 0.1 for $n=30$. This finding highlights the fact that any particular subjectively elicited value---e.g., the widely-used value of 0.04 for the natural conjugate prior; see \citet{CCM16} and \citet{chan20}---is unlikely to be suitable for all datasets of different dimensions. In addition, the estimates also suggest that when the model dimension increases, more aggressive shrinkage is needed, confirming the findings in, e.g., \citet{BGR10}. 

Second, it is clear that the estimates of $\kappa_1$ are orders of magnitude larger than those of $\kappa_2$ under VAR-SV and VAR-FSV, the two models that accommodate cross-variable shrinkage. For example, for VAR-FSV with $n=15$, the estimates of $\kappa_1$ and $\kappa_2$ are, respectively, 0.17 and 0.0018, a difference of 2 orders of magnitude. These estimates confirm the model comparison results above and suggest that the data strongly prefers shrinking the coefficients on lags of other variables much more aggressively to zero than those on own lags. Again, this is intuitive as one would expect that own lags are more informative, on average, than lags of other variables in forecasting a variable's future evolution. In addition, these estimates are also rather different from those widely-used subjectively chosen values of $\kappa_1 = 0.04$ and $\kappa_2 = 0.0016$, reinforcing the conclusion that any particular set of fixed hyperparameter values is unlikely to be 
suitable for all datasets of different dimensions. 

\begin{table}[H]
\caption{Posterior means (standard deviations) of the shrinkage hyperparameters in
VAR-CSV, VAR-SV and VAR-FSV.}
\label{tab:kappas}
\centering
\begin{tabular}{lccccc}
\hline\hline
	&	VAR-CSV	&	\multicolumn{2}{c}{VAR-SV}			&	\multicolumn{2}{c}{VAR-FSV} \\ 
	&	$\kappa$	&	$\kappa_1$	&	$\kappa_2$	&	$\kappa_1$	&	$\kappa_2$	\\ \hline
$n=7$	&	0.32	&	0.23	&	0.0032  	&	0.24	&	0.0037	\\
	&	(0.065)	&	(0.051)	&	(0.0014)	&	(0.054)	&	(0.0015)	\\
	\rowcolor{lightgray}
$n=15$	&	0.22	&	0.16	&	0.0022	&	0.17	&	0.0018	\\
	\rowcolor{lightgray}
	&	(0.027)	&	(0.030)	&	(0.0006)	&	(0.032)	&	(0.0004)	\\
$n=30$	&	0.10	&	0.15	&	0.00002	&	0.15	&	0.00003	\\
	&	(0.008)	&	(0.020)	&	(0.00001)	&	(0.020)	&	(0.00001)	\\ \hline\hline
\end{tabular}
\end{table}

All in all, our results confirm the substantial benefits of allowing for cross-variable shrinkage and selecting the shrinkage hyperparameters using a a data-based approach. They also highlight that in high-dimensional settings, choosing a flexible shrinkage prior is as important as selecting a flexible stochastic volatility specification. 

\section{Extensions} \label{s:extensions}

The proposed marginal likelihood estimators are based on two key ingredients: a conditional Monte Carlo component to integrate out the large number of VAR coefficients analytically and an adaptive importance sampling method to integrate out the log-volatility via Monte Carlo simulation. These two ideas can be more widely applied to other marginal likelihood estimators or alternative model comparison methods to make them viable in high-dimensional settings, as we demonstrate below. In addition, we further illustrate how the proposed methods can be used to compute the marginal likelihood of a stochastic volatility model with an outlier component.

\subsection{Improving the Modified Harmonic Mean Estimator}

We first illustrate how one can use the conditional Monte Carlo and adaptive importance sampling methods to improve alternative estimators of the marginal likelihood. To fix ideas, we focus on the modified harmonic mean estimator proposed by \citet{GD94}; similar improvements can be made for other estimators such as Chib's method \citep{chib95, CJ01} and sequential Monte Carlo \citep[e.g.,][]{BZ20}.

The theoretical justification of the modified harmonic mean estimator is based on the following observation: for any density function $f$ with support contained in the support of the posterior density, we have
\begin{equation} \label{eq:GD_proof}
	\Em\left( \frac{f(\vect{\theta})}{p(\vect{\theta})p(\by\gvn \vect{\theta})}\;\bigg|\; \by\right)
	= \int\frac{f(\vect{\theta})}{p(\vect{\theta})p(\by\gvn \vect{\theta})}
	\frac{p(\vect{\theta})p(\by\gvn\vect{\theta})}{p(\by)}\, \di\vect{\theta}	=	p(\by)^{-1},
\end{equation}
where the expectation is taken with respect to the posterior distribution $p(\vtheta\gvn\by)=p(\vect{\theta})p(\by\gvn \vect{\theta})/p(\by)$. It follows that one can estimate the marginal likelihood $p(\by)$ using the following estimator:
\begin{equation}\label{eq:GD}
    \text{GD} =
    \left\{
        \frac{1}{R}\sum_{r=1}^{R}\frac{f(\vect{\theta}^{(r)})}{p(\vect{\theta}^{(r)})p(\by\gvn \vect{\theta}^{(r)})}
    \right\}^{-1},
\end{equation}
where $\vtheta^{(1)},\ldots, \vtheta^{(R)}$ are posterior draws. This modified harmonic mean is simulation-consistent---it converges to $p(\by)$ in probability as $R$ tends to infinity---but it is not unbiased in general. 

Even though the estimator in \eqref{eq:GD} is simulation-consistent for any density $f$ with support contained in the support of $p(\vtheta\gvn\by)$, its finite-sample properties depend critically on the choice of $f$. \citet{Geweke99} shows that if $f$ has tails lighter than those of $p(\vtheta\gvn\by)$, the estimator has a finite variance. He further recommends a Gaussian approximation of $p(\vtheta\gvn\by)$ with tail truncations determined by asymptotic arguments. More specifically, let $\hat{\vtheta}$ and $\mathbf{Q}_{\vtheta}$ denote the posterior mean and covariance matrix respectively. Then, $f$ is set to be the
$\distn{N}(\hat{\vtheta}, \mathbf{Q}_{\vtheta})$ density truncated within the region 
\[
	\{\vtheta\in\mathbb{R}^m: (\vtheta-\hat{\vtheta})'\mathbf{Q}_{\vtheta}^{-1}(\vtheta-\hat{\vtheta})<\chi^2_{\alpha,m}\},
\]
where $\chi^2_{\alpha,m}$ is the $(1-\alpha)$ quantile of the $\chi^2_m$ distribution and $m$ is the dimension of $\vtheta$. The $\text{GD}$ estimator in \eqref{eq:GD} with the truncated Gaussian tuning function is commonly-used in economics; examples include \citet{SWZ08} and \citet{JP08}.

One potential difficulty of implementing this approach in high-dimensional setting is that the prescribed Gaussian density has a large number of parameters to be estimated. As an example, for the VAR-SV model with $T=300, n = 20,$ and $p=4$,  the Gaussian density is 2,133-dimensional and has over 2.2 million parameters. As such, it is impractical to estimate them with any reasonable accuracy using MCMC output.

Below we outline a few ways to make the $\text{GD}$ estimator viable in high-dimensional settings using the proposed adaptive importance sampling and conditional Monte Carlo methods. First, we consider a tuning function of the form $f(\vtheta) = f_{\bh}(\bh)f_{\vtheta_{-\bh}}(\vtheta_{-\bh})$, where $\vtheta_{-\bh}$ denotes the set of model parameters and latent variables excluding the log-volatility $\bh$. Then, we use the adaptive importance sampling method described in Section~\ref{ss:IS} to obtain $f_{\bh}$, whereas $f_{\vtheta_{-\bh}}$ is a truncated Gaussian distribution constructed as described above. This approach reduces the number of parameters of $f$ in two ways: 1) by factoring $f$ into two lower-dimensional Gaussian densities, it reduces the number of covariance parameters; 2) as described in Section~\ref{ss:IS}, the way $f_{\bh}$ is parameterized reduces the number of parameters from $(T^2+3T)/2$ to $2T+1$.\footnote{As discussed in Section~\ref{ss:CSV}, the choice of the restricted family of Gaussian densities exploits the time-series structure of the problem. In particular, the log-volatility is assumed to follow an AR(1) process, which implies the precision or inverse covariance matrix of the joint distribution is tridiagonal. This strikes the balance between flexibility and parsimony.} However, this approach alone is unlikely to work well as there remains a large number of parameters to be estimated. Using the same example of the VAR-SV with $T=300, n = 20,$ and $p=4$, the tuning function has still over 1.6 million parameters.

Next, we incorporate the proposed conditional Monte Carlo estimator to improve the $\text{GD}$ estimator. Using the notations in Section~\ref{ss:CMC}, partition $\vtheta$ into two blocks, $\vgamma$ and $\vpsi$, and suppose we can compute $g(\by\gvn\vpsi) \equiv \Em[p(\by\gvn\vtheta)\gvn \vpsi] = \int p(\by\gvn\vgamma,\vpsi)p(\vgamma\gvn\vpsi)\di \vgamma$ analytically, where $p(\vgamma\gvn \vpsi)$ is the prior of $\vgamma$ that can potentially depend on $\vpsi$. Since the marginal likelihood can be expressed as $p(\by) = \int g(\by\gvn\vpsi)p(\vpsi)\di\vpsi$, a similar derivation as \eqref{eq:GD_proof} suggests that we can estimate $p(\by)$ using the following estimator:
\begin{equation}\label{eq:GD_psi}
    \text{GD}_{\vpsi} =
    \left\{
        \frac{1}{R}\sum_{r=1}^{R}\frac{f_{\vpsi}(\vpsi^{(r)})}
				{p(\vpsi^{(r)})g(\by\gvn \vpsi^{(r)})}
    \right\}^{-1},
\end{equation}
where $\vpsi^{(1)},\ldots, \vpsi^{(R)}$ are posterior draws from $p(\vpsi\gvn\by) = 
g(\by\gvn\vpsi)p(\vpsi)/p(\by)$. These draws can be obtained using a standard MCMC algorithm: first obtain posterior draws of the pairs $(\vgamma^{(1)},\vpsi^{(1)}),\ldots, (\vgamma^{(R)},\vpsi^{(R)})$ and then discard the first components. One can use the truncated Gaussian density described above as the tuning function $f_{\vpsi}$. Using the same VAR-SV example, the number of parameters of the tuning function here is drastically reduced to about 46,000. Even so, estimating these parameters with sufficient accuracy using MCMC output is still computationally intensive.

Finally, one can combine both approaches to further improve the performance of the estimator. Specifically, consider the estimator in \eqref{eq:GD_psi} and a tuning function of the form $f_{\vpsi}(\vpsi) = f_{\bh}(\bh)f_{\vpsi_{-\bh}}(\vpsi_{-\bh})$, where $ f_{\bh}$ is obtained using the adaptive importance sampling method and $f_{\vpsi_{-\bh}}$ is a truncated Gaussian density. For the same VAR-SV example, the number of parameters for this combined approach is only 610, and they can be estimated with adequate accuracy using MCMC output.

\begin{figure}[H]
  \centering
  \includegraphics[width=.6\textwidth]{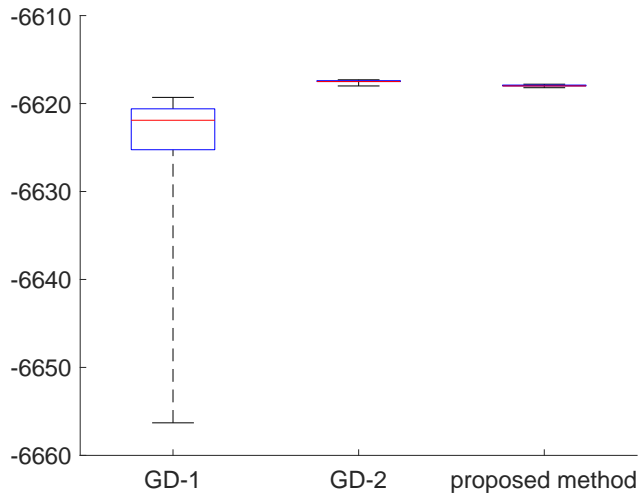}
  \caption{Boxplots of the log marginal likelihoods for VAR-CSV with $n=15$ estimated using three estimators: the GD estimator in which $\vgamma = \{\bA,\vSigma,\sigma^2\}$ is integrated out analytically (GD-1), the GD estimator in which $\vgamma$ is integrated out analytically and the adaptive importance sampling method is used to obtain $f_{\bh}$ (GD-2) and the proposed method.}
   \label{fig:GD}
\end{figure}

To illustrate how these improved estimators perform empirically, we revisit the model comparison exercise in Section~\ref{s:application} and compute the marginal likelihood of VAR-CSV with $n=15$ using two versions of the GD estimator. Recall that the set of model parameters and latent variables for VAR-CSV is $\vtheta = \{\bA, \vSigma, \phi, \sigma^2, \kappa, \bh\}$. We first consider the estimator in~\eqref{eq:GD_psi}, where we integrate out $\vgamma = \{\bA,\vSigma,\sigma^2\}$ analytically and set $f_{\vpsi}$ to be a truncated Gaussian density with $\vpsi = \{\phi, \kappa, \bh\}$. Second, we use the same estimator in~\eqref{eq:GD_psi}, but with a different tuning function: we set $f_{\vpsi}(\vpsi) = f_{\bh}(\bh)f_{\vpsi_{-\bh}}(\vpsi_{-\bh})$, where $ f_{\bh}$ is obtained using the adaptive importance sampling method and $f_{\vpsi_{-\bh}}$ is a truncated bivariate Gaussian density with $\vpsi_{-\bh} = \{\phi, \kappa\}$. For each estimator, we obtain 100 marginal likelihood estimates, where each estimate is based on 50,000 posterior draws. To summarize the results we report the boxplots of the log marginal likelihood estimates in Figure~\ref{fig:GD}. The middle line of each box denotes the median, while the lower and upper lines represent, respectively, the 25- and the 75-percentiles. The whiskers extend to the maximum and minimum. For comparison we also include the estimates from the proposed method.\footnote{The original GD estimator in \eqref{eq:GD} and the version with only the adaptive importance sampling method have too many parameters, and they are unstable in this high-dimensional setting.}

While the original GD estimator in \eqref{eq:GD} has too many parameters to provide stable estimates, the version in which $\vgamma = \{\bA,\vSigma,\sigma^2\}$ is integrated out analytically (GD-1) performs reasonably well. In particular, the variance of the estimator is relatively small and most of the marginal likelihood estimates are between $-6,630$ and $-6,620$. Nevertheless, the distribution of the estimator has a heavier left tail and it seems to underestimate the value of the marginal likelihood. In contrast, when both the conditional Monte Carlo and adaptive importance sampling methods are used to improve the GD estimator (GD-2), its variance is substantially reduced and its performance is virtually the same as the proposed estimator that is based on the cross-entropy method. In particular, all estimates of both estimators are around $-6,618$.

\subsection{The Observe-Data Deviance Information Criterion}

Another commonly-used model comparison criterion is the deviance information criterion (DIC) introduced in the seminal paper by \citet{SBC02}. Below we discuss how one can use the proposed conditional Monte Carlo and adaptive importance sampling methods to efficiently compute the DIC for high-dimensional stochastic volatility models.

The DIC is based on the concept of deviance, which is defined as
\[
	D(\vtheta) = -2\log p(\by\gvn\vtheta),
\]
where $p(\by\gvn\vtheta)$ is the (observed-data) likelihood.\footnote{Here we distinguish between the conditional likelihood, i.e., the data density conditional on the latent variables, and the observed-data likelihood (data density marginal of the latent variables). While the DIC can be defined in terms of the conditional likelihood \citep{CFRT06}, a few recent papers have pointed out that this version is problematic for theoretical and practical reasons \citep[see, e.g.,][]{CG16, LYZ20}. We therefore focus on the original DIC based on the observed-data likelihood.} Next, the effective number of parameters, $p_{D}$, of the model is defined to be
\[
	p_{D} = \overline{D(\vtheta)} - D(\widetilde{\vtheta}),
\]
where $	\overline{D(\vtheta)} = -2 \Em_{\vtheta}[ \log p(\by\gvn\vtheta) \gvn \by]$ is the posterior mean deviance and $\widetilde{\vtheta}$ is an estimate of $\vtheta$, which is typically taken as the
posterior mean or mode. Finally, the deviance information criterion is the sum of the posterior mean deviance, which can be viewed as a Bayesian measure of model fit, and the effective number of parameters that measures model complexity:
\[
	\text{DIC} = \overline{D(\vtheta)} + p_{D}.
\]
The DIC may therefore be interpreted as a trade-off between model fit and model complexity. Given a set of competing models, the preferred model is the one with the minimum DIC value. 

The main difficulty in computing the DIC for high-dimensional stochastic volatility models is that the likelihood is not available in closed-form and it is typically computationally intensive to evaluate. We show how one can use the conditional Monte Carlo and adaptive importance sampling methods developed in Section~\ref{s:ML} to overcome this difficulty. To fix ideas, we focus on the likelihood evaluation of the VAR-FSV model; likelihood evaluation for other stochastic volatility models can be done similarly.

The latent variables in the VAR-FSV model are the factors and the log-volatility. We aim to integrate out the factors analytically and the log-volatility using importance sampling. To that end, we stack the observations $\by = (\by_1',\ldots, \by_T')'$ over $t=1,\ldots, T$ and write the VAR-FSV model as
\[
	\by = \bX\valpha + (\mathbf{I}_T\otimes\bL)\bbf, \quad \bu^y\sim\distn{N}(\mathbf{0},\bD), \quad \bbf \sim\distn{N}(\mathbf{0},\bG),
\]
where $\bX$ is an appropriately defined design matrix consisting of intercepts and lagged values, $\bD = \text{diag}(\bD_1,\ldots, \bD_T)$ and $\bG = \text{diag}(\bG_1,\ldots, \bG_T)$ with $\bD_t = \text{diag}(\e^{h_{1,t}},\ldots, \e^{h_{n,t}})$ and $\bG_t = \text{diag}(\e^{h_{n+1,t}},\ldots, \e^{h_{n+r,t}})$. Hence, the distribution of $\by$ marginal of the factors $\bbf$ is $(\by\gvn \valpha, \bl,\bh) \sim \distn{N}(\bX\valpha,\bS_{y})$ with density function
\[
	p(\by\gvn \valpha, \bl,\bh) = (2\pi)^{-\frac{Tn}{2}}|\bS_y|^{-\frac{1}{2}}
		\e^{-\frac{1}{2}(\by-\bX\valpha)'\bS_y^{-1}(\by-\bX\valpha)},
\]
where $\bS_{y} = \bD + (\mathbf{I}_T\otimes\bL)\bG(\mathbf{I}_T\otimes\bL')$. Finally, the likelihood function (marginal of both $\bbf$ and $\bh$ but conditional on the time-invariant model parameters) can be evaluated via:
\begin{align*}
	p(\by\gvn \valpha, \bl,\vmu,\vphi,\vsigma^2) = \int p(\by\gvn \valpha, \bl,\bh) \prod_{j=1}^{n+r}p(\bh_{j,\bigcdot}\gvn\mu_j,\phi_j,\sigma_j^2)\di \bh,
\end{align*}
where $\bh_{j,\bigcdot} = (h_{j,1},\ldots, h_{j,T})'$ and $p(\bh_{j,\bigcdot}\gvn\mu_j,\phi_j,\sigma_j^2)$ is a Gaussian density that is available analytically. To implement the importance sampling approach described in Section~\ref{ss:IS}, we consider the parametric family
\[
	\mathcal{F} = \left\{\prod_{j=1}^{n+r}f_{\distn{N}}(\bh_{j,\bigcdot}; \hat{\bh}_{j,\bigcdot}, \hat{\bK}_{\bh_{j,\bigcdot}}^{-1})\right\}.
\]
The parameters of the importance sampling densities are chosen by solving the maximization problem in~\eqref{eq:maxMC}. In particular, all the $T$-variate Gaussian densities are obtained using the procedure described in Section~\ref{ss:CSV}.

To illustrate the methods, we compute the observed-data DIC of the 3 stochastic volatility models with $n=7$, and the results are reported in Table~\ref{tab:DIC}. In contrast to the marginal likelihood results, VAR-CSV is the most favored model according to the observed-data DIC. This is due to a combination of better model-fit (as shown by a lower posterior mean deviance value) and lower model complexity (in terms of a lower effective number of parameters). Despite being commonly used in the literature, rigorous justification of the DIC as a model selection criterion has only been recently investigated \citep{LYZ21}. More research to better understand the theoretical and empirical differences between the DIC and the marginal likelihood is therefore needed.

\begin{table}[H]
\caption{The observed-data DIC, the posterior mean deviance $\overline{D(\vtheta)}$	and the effective number of parameters $p_{D}$ of VAR-CSV, VAR-SV and VAR-FSV for $n=7$.}
\label{tab:DIC}
\centering
\begin{tabular}{lccc}
\hline\hline
	  &	VAR-CSV	&	VAR-SV	&	VAR-FSV	\\ \hline
DIC	&	3,367	&	3,440	&	3,517	\\
\rowcolor{lightgray}
$ \overline{D(\vtheta)}$	&	3,181	&	3,195	&	3,239	\\
$p_{D}$	&	186	&	245	&	278	\\
 \hline\hline
\end{tabular}
\end{table}

\subsection{A Stochastic Volatility Model with an Outlier Component}

There has been a lot of recent work to develop VARs that can handle COVID-19 outliers; examples include \citet{Hartwig21}, \citet{Ng21}, \citet{LP22} and \citet{CCMM22}. In particular, \citet{CCMM22} propose various stochastic volatility specifications that have an explicit component to model infrequent volatility outliers. Below we consider one of these stochastic volatility models. It extends the Cholesky stochastic volatility by incorporating a discrete mixture representation proposed in \citet{SW16} for handling outliers in unobserved component models. More specifically, with the same reduced-form VAR in \eqref{eq:VAR}, the innovation $\vepsilon_t$ is now distributed as:
\begin{equation} \label{eq:VAR-SVO}
	\vepsilon_t \sim\distn{N}(\mathbf{0}, \vSigma_t), \quad \vSigma_t^{-1} = \bB_0' \bD_t^{-1}\bB_0 o_t^{-2}.
\end{equation}
The new addition is the outlier parameter $o_t$, which is modeled to have a mixture distribution that distinguishes between regular observations with $o_t=1$ and outliers for which $o_t>2$. More precisely, with probability $1-p_o$, $o_t=1$; and with probability $p_o$, $o_t$ follows a uniform distribution on $(2,20)$: $o_t\sim\mathcal{U}(2,20)$. The outlier probability $p_o$ is assumed to have a beta prior, where its hyperparameters are calibrated so that the mean outlier frequency is once every 4 years in quarterly data. We refer to this model as VAR-SVO.

Following \citet{SW16} and \citet{CCMM22}, the distribution of $o_t$ is discretized using a grid. Consequently, $o_t$ follows a discrete distribution that can be easily handled. We refer the readers 
to \citet{CCMM22} for estimation details. We fit VAR-SVO using the datasets ($n=7, 15, 30$) in Section~\ref{s:application} and compute the corresponding marginal likelihood values. The results are reported in Table~\ref{tab:ML3}. For convenience, we also reproduce the marginal likelihood values of other models (the value of the best VAR-FSV for each dimension is reported).

\begin{table}[H]
\caption{Log marginal likelihood estimates (numerical standard errors) of a standard homoskedastic VAR, VAR-CSV, VAR-SV, VAR-FSV and VAR-SVO.}
\label{tab:ML3}
\centering
\begin{tabular}{lccccc}
\hline\hline
	&	VAR	&	VAR-CSV	&	VAR-SV	&	VAR-FSV	&	VAR-SVO	\\ \hline
$n=7$	&	$-$2,583.4	&	$-$2,410	&	$-$2,315	&	$-$2,318	&	$-$2,306	\\
	&	--	&	(0.1)	&	(0.3)	&	(0.4)	&	(0.6)	\\
	\rowcolor{lightgray}
$n=15$	&	$-$6,918.8	&	$-$6,618	&	$-$6,443	&	$-$6,454	&	$-$6,386	\\
\rowcolor{lightgray}
	&	--	&	(0.1)	&	(0.7)	&	(0.8)	&	(0.8)	\\
$n=30$	&	$-$12,783.4	&	$-$12,024	&	$-$11,382	&	$-$11,567	&	$-$11,255	\\
	&	--	&	(0.1)	&	(1.2)	&	(1.8)	&	(1.9)	\\  \hline\hline
\end{tabular}
\end{table}

For all model dimensions, VAR-SVO is the most favored specification by the data. These results suggest that there are infrequent, extreme observations even before the COVID-19 pandemic. Moreover, they also highlight that an explicit outlier component is useful for handling these apparent outliers, even in the presence of time-varying volatility.

\section{Concluding Remarks and Future Research} \label{s:conclusion}

As large Bayesian VARs are now widely used in empirical applications, choosing the most suitable stochastic volatility specification and shrinkage priors for a particular dataset has become an increasingly pertinent problem. We took a first step to address this issue by developing Bayesian model comparison methods to select among a variety of stochastic volatility specifications and shrinkage priors in the context of large VARs. We demonstrated via a series of Monte Carlo experiments that the proposed method worked well---particularly it could be used to discriminate VARs with different stochastic volatility specifications. 

Using US datasets of different dimensions, we showed that the data strongly preferred the Cholesky stochastic volatility, whereas the factor stochastic volatility was also competitive. This finding thus suggests that more future research on factor stochastic volatility models would be fruitful, given that they are not as commonly used in empirical macroeconomics. Our results also confirmed the vital role of flexible shrinkage priors: both cross-variable shrinkage and a data-based approach to determine the overall shrinkage strength were empirically important. 

In future work, it would be useful to extend the proposed estimators to compare large time-varying parameter VARs. Existing evidence seems to suggest that in a large VAR, only a few of the coefficients are time-varying. Such a model comparison method would be helpful for comparing different dynamic shrinkage priors, and thus provide useful guidelines for empirical researchers.

\newpage

\section*{Appendix A: Estimation Details}

In this appendix we provide the details of the priors and the estimation of the Bayesian VARs with three different stochastic volatility specifications. We also discuss the details of the adaptive importance sampling approach for estimating the marginal likelihoods for these models.  

\subsection*{Common Stochastic Volatility}

We first outline the estimation of the common stochastic volatility model given in \eqref{eq:VAR}--\eqref{eq:h}. To that end, let $\bx_t' = (1, \by_{t-1}',\ldots, \by_{t-p}')$ be a $1\times k$ vector of an intercept and lags with $k=1+np$. Then, stacking the observations over $t=1,\ldots, T$, we have
\[
	\bY = \bX\bA + \vepsilon, 
\]
where $\bA = (\ba_0, \bA_1, \ldots, \bA_p)'$ is $k\times n$, and the matrices $\bY$, $\bX$ and 
$\vepsilon$ are respectively of dimensions $T\times n$, $T\times k$ and $T\times n$. Under the common stochastic volatility model, the innovations are distributed as $\text{vec}(\vepsilon)\sim\distn{N}(\mathbf{0},\vSigma\otimes \bD_{\bh})$, where $\bD_{\bh} = \text{diag}(\e^{h_1},\ldots, \e^{h_T})$. The log-volatility $h_t$ evolves as a stationary AR(1) process given in \eqref{eq:h}, 
which for convenience we reproduce below:
\[
	h_t = \phi h_{t-1} + u_t^h, \quad u_t^h\sim\distn{N}(0,\sigma^2), 
\]
for $t=2,\ldots, T$, where the process is initialized as $h_{1}\sim\distn{N}(0,\sigma^2/(1-\phi^2))$.

Next, we describe the priors on the model parameters  $\bA, \vSigma, \kappa, \phi$ and $\sigma^2$. First, we consider a natural conjugate prior on $(\bA, \vSigma \gvn \kappa)$:
\[
	\vSigma\sim\distn{IW}(\nu_0,\bS_0), \quad (\text{vec}(\bA)\gvn\vSigma,\kappa) 
		\sim\distn{N}(\text{vec}(\bA_0), \vSigma\otimes \bV_{\bA}),
\]	
where the prior hyperparameters $\text{vec}(\bA_0)$ and $\bV_{\bA}$ are chosen in the spirit of the Minnesota prior. More specifically, we set $\bA_0 = \mathbf{0}$ and the covariance matrix $\bV_{\bA}$ is assumed to be diagonal with diagonal elements $v_{\bA,ii} = \kappa/(l^2 \hat{s}_r)$ for a coefficient associated to the $l$-th lag of variable $r$ and $v_{\bA,ii} = 100$ for an intercept, where $\hat{s}_r$ is the sample variance of an AR(4) model for the variable $r$. Note that here a single hyperparameter $\kappa$ controls the overall shrinkage strength and this prior does not distinguish `own' versus `other' lags. Here we treat $\kappa$ to be an unknown parameter with a hierarchical gamma prior: $\kappa\sim\distn{G}(c_{1},c_{2})$. Finally, for the prior distributions of $\phi$ and $\sigma^2$, they are respectively truncated normal and inverse-gamma: 
$\phi\sim\distn{N}(\phi_0,V_{\phi})1(|\phi|<1)$ and $\sigma^2 \sim\distn{IG}(\nu_{\sigma^2},S_{\sigma^2})$, where $1(\cdot)$ denotes the indicator function.

With the priors on the model parameters specified above, one can obtain posterior draws by sequentially sampling from:
\begin{enumerate}
	\item $p(\bA, \vSigma \gvn \bY, \bh, \phi,\sigma^2,\kappa) = p(\bA, \vSigma \gvn \bY, \bh, \kappa)$;
	\item $p(\bh \gvn \bY, \bA, \vSigma, \phi,\sigma^2,\kappa) = p(\bh \gvn \bY, \bA, \vSigma, \phi, \sigma^2)$;
	\item $p(\phi \gvn \bY, \bA, \vSigma, \bh, \sigma^2,\kappa) = p(\phi \gvn \bh, \sigma^2)$;
	\item $p(\sigma^2 \gvn \bY, \bA, \vSigma, \bh, \phi, \kappa) = p(\sigma^2 \gvn \bh, \phi)$; 
	\item $p(\kappa \gvn \bY,\bA, \vSigma, \bh, \phi,\sigma^2) = p(\kappa \gvn \bA,\vSigma)$. 
\end{enumerate}

\textbf{Step~1}: we use the results proved in \citet{chan20} that $(\bA, \vSigma \gvn \bY, \bh,\kappa)$ has a normal-inverse-Wishart distribution. More specifically,  let 
\begin{align*}
	\bK_{\bA} & = \bV_{\bA}^{-1} + \bX'\bD^{-1}_{\bh} \bX, \\
	\hat{\bA} & = \bK_{\bA}^{-1}(\bV_{\bA}^{-1}\bA_0 +  \bX'\bD^{-1}_{\bh} \bY), \\
	\hat{\bS} & = \bS_0 + \bA_0'\bV_{\bA}^{-1}\bA_0 + \bY'\bD^{-1}_{\bh}\bY
	-\hat{\bA}'\bK_{\bA}\hat{\bA}.
\end{align*}
Then $(\bA, \vSigma \gvn \bY, \bh,\kappa)$ has a normal-inverse-Wishart distribution with parameters $\nu_0+T$, $\hat{\bS}$, $\hat{\bA}$ and $\bK_{\bA}^{-1}$. We can sample $(\bA, \vSigma \gvn \bY, \bh, \kappa)$ in two steps. First, sample $\vSigma $ marginally from the inverse-Wishart distribution $(\vSigma \gvn \bY, \bh) \sim\distn{IW}(\nu_0+T,\hat{\bS})$. Then, given the sampled $\vSigma$, obtain $\bA$ from the normal distribution:
\[
	(\text{vec}(\bA)\gvn \bY,\vSigma,\bh) \sim \distn{N}(\text{vec}(\hat{\bA}), 
	\vSigma\otimes \bK_{\bA}^{-1}).
\]
We note that one can sample from this normal distribution efficiently without explicitly computing the inverse $\bK_{\bA}^{-1}$; we  refer the readers to \citet{chan20} for computational details.

\textbf{Step 2}: note that
\[
	p(\bh \gvn \bY, \bA, \vSigma, \phi,\sigma^2) \propto p(\bh\gvn\phi,\sigma^2) 
	\prod_{t=1}^T	p(\by_t\gvn\bA, \vSigma,h_t),
\]
where $p(\bh\gvn\phi,\sigma^2)$ is a Gaussian density implied by the state equation 
\eqref{eq:h}. The log conditional likelihood $p(\by_t\gvn\bA, \vSigma,h_t)$ has the following explicit expression:
\[
	\log p(\by_t\gvn\bA, \vSigma, h_t) = c_t -\frac{n}{2}h_t - \frac{1}{2}\e^{-h_t}\vepsilon_t'\vSigma^{-1}\vepsilon_t,
\]
where $c_t$ is a constant independent of $h_t$. It is easy to check that
\begin{align*}
	\frac{\partial}{\partial h_t}\log p(\by_t\gvn\bA, \vSigma, h_t) & = -\frac{n}{2} + \frac{1}{2}\e^{-h_t}\vepsilon_t'\vSigma^{-1}\vepsilon_t, \\
	\frac{\partial^2}{\partial h_t^2}\log p(\by_t\gvn\bA, \vSigma, h_t) & 
	=- \frac{1}{2}\e^{-h_t}\vepsilon_t'\vSigma^{-1}\vepsilon_t < 0.
\end{align*}
Given the above first and second derivatives of the log conditional likelihood, one can use the Newton-Raphson algorithm to obtain the mode of $\log p(\bh \gvn \bY, \bA, \vSigma, \phi,\sigma^2)$ and compute the negative Hessian evaluated at the mode, which are denoted as $\hat{\bh}$ and $\bK_{\bh}$, respectively. Since the Hessian is negative definite everywhere, 
$\bK_{\bh}$ is positive definite. Next, using $\distn{N}(\hat{\bh}, \bK_{\bh}^{-1})$ as a proposal distribution, one can sample $\bh$ directly using an acceptance-rejection Metropolis-Hastings step. We refer the readers to \citet{chan20} for details. 

\textbf{Step 3}: the conditional density of $\phi$ is of the form
\[
	p(\phi \gvn \bh, \sigma^2)\propto p(\phi)g(\phi)\e^{-\frac{1}{2\sigma^2}\sum_{t=2}^T(h_t-\phi h_{t-1})^2},
\]
where $ g(\phi) = (1-\phi^2)^{\frac{1}{2}}\e^{-\frac{1}{2\sigma^2}(1-\phi^2)h_1^2}$ and $p(\phi)$ is the truncated normal prior. This density is nonstandard, but we can sample $\phi$ via an independence-chain Metropolis-Hastings step using the proposal distribution $\distn{N}(\hat{\phi}, K_{\phi}^{-1}) 1(|\phi|<1)$, where 
\[
	K_{\phi}     = V_{\phi}^{-1} + \frac{1}{\sigma^2}\sum_{t=2}^{T}h_{t-1}^2 \quad 
	\hat{\phi} = K_{\phi}^{-1}\left(V_{\phi}^{-1}\phi_{0} + \frac{1}{\sigma^2}\sum_{t=2}^{T}h_th_{t-1}\right).
\]
Then, given the current draw $\phi$, a proposal $\phi^*$ is accepted with probability $\min(1,g(\phi^*)/g(\phi))$; otherwise the Markov chain stays at the current state $\phi$.

\textbf{Step 4}: $\sigma^2$ can be sampled easily as it has the following inverse-gamma distribution:
\[
	(\sigma^2 \gvn \bh, \phi) \sim \distn{IG}(\nu_{\sigma^2}+T/2, \widetilde{S}_{\sigma^2}),
\]
where $ \widetilde{S}_{\sigma^2} = S_{\sigma^2} + \left[(1-\phi^2)h_1^2 + \sum_{t=2}^T(h_{t}-\phi h_{t-1})^2\right]/2$.

\textbf{Step 5}: first note that $\kappa$ only appears in its gamma prior $\kappa\sim \distn{G}(c_{1},c_{2})$ and $\bV_{\bA}$, the prior covariance matrix of $\bA$. Recall that $\bV_{\bA}$ is a $k\times k$ diagonal matrix in which the first element---corresponding to the prior variance of the intercept---does not involve~$\kappa$. More explicitly, for $i=2,\ldots, k$, write the $i$-th diagonal element of $\bV_{\bA}$ as 
$v_{\bA,ii} = \kappa C_{i}$ for some constant $C_i$. Then, we have
\begin{align*}
	p(\kappa \gvn \bA,\vSigma) & \propto \kappa^{c_1-1}\e^{-c_2\kappa} \times |\bV_{\bA}|^{-\frac{n}{2}} \e^{-\frac{1}{2}\text{tr}\left(\vSigma^{-1}(\bA-\bA_0)'\bV_{\bA}^{-1}(\bA-\bA_0)\right)} \\	
	& \propto \kappa^{c_1-\frac{(k-1)n}{2}-1}\e^{-c_2\kappa} 
	\e^{-\frac{1}{2}\text{tr}\left(\bV_{\bA}^{-1}(\bA-\bA_0)\vSigma^{-1}(\bA-\bA_0)'\right)} \\
	 & \propto \kappa^{c_1-\frac{(k-1)n}{2}-1} \e^{-\frac{1}{2} \left(2c_2\kappa + \kappa^{-1}
	\sum_{i=2}^k Q_i/C_i\right)},
\end{align*}
where $Q_i$ is the $i$-th diagonal element of $\bQ = (\bA-\bA_0)\vSigma^{-1}(\bA-\bA_0)'$.
Note that this is the kernel of the generalized inverse Gaussian distribution $\distn{GIG}\left(c_{1}-\frac{(k-1)n}{2}, 2c_{2}, \sum_{i=2}^k Q_i/C_i\right)$. Draws from the  generalized inverse Gaussian distribution can be obtained using the algorithm in \citet{devroye2014}.

Next, we derive the expression of $p(\bY\gvn\bh,\kappa)$ given in \eqref{eq:ygvnh}. First let 
$k_1$ denote the normalizing constant of the normal-inverse-Wishart prior: 
 $k_1 = (2\pi)^{-\frac{nk}{2}}2^{-\frac{n\nu_0}{2}}|\bV_{\bA}|^{-\frac{n}{2}}
\Gamma_n(\nu_0/2)^{-1}|\bS_0|^{\frac{\nu_0}{2}}$. Then, by direct computation:
\begin{align*}
	p(\bY\gvn\bh,\kappa) & = \int p(\bY\gvn\bA,\vSigma,\bh)p(\bA,\vSigma\gvn \kappa) 
	\di(\bA,\vSigma)\\
	& = \int (2\pi)^{-\frac{Tn}{2}}|\vSigma|^{-\frac{T}{2}}\e^{-\frac{n}{2}\mathbf{1}_T'\bh}
	\e^{-\frac{1}{2}\text{tr}\left(\vSigma^{-1}(\bY-\bX\bA)'\bD^{-1}_{\bh}(\bY-\bX\bA)\right)}\\
	& \quad \times k_1 |\vSigma|^{-\frac{\nu_0+n+k+1}{2}}\e^{-\frac{1}{2}\text{tr}\left(\vSigma^{-1}\bS_0\right)}
	\e^{-\frac{1}{2}\text{tr}\left(\vSigma^{-1}(\bA-\bA_0)'\bV_{\bA}^{-1}(\bA-\bA_0)\right)} \di(\bA,\vSigma) \\	
	& = k_1(2\pi)^{-\frac{Tn}{2}}\e^{-\frac{n}{2}\mathbf{1}_T'\bh}\int
		|\vSigma|^{-\frac{\nu_0+T+n+k+1}{2}}\e^{-\frac{1}{2}\text{tr}\left(\vSigma^{-1}\hat{\bS}\right)}
			\e^{-\frac{1}{2}\text{tr}\left(\vSigma^{-1}(\bA-\hat{\bA})'\bK_{\bA}^{-1}(\bA-\hat{\bA})\right)}	
	 \di(\bA,\vSigma) \\	
	& = \pi^{-\frac{Tn}{2}}\e^{-\frac{n}{2}\mathbf{1}_T'\bh}
	|\bV_{\bA}|^{-\frac{n}{2}}|\bK_{\bA}|^{-\frac{n}{2}}
	\frac{\Gamma_n\left(\frac{\nu_0+T}{2}\right)|\bS_0|^{\frac{\nu_0}{2}}}
	{\Gamma_n\left(\frac{\nu_0}{2}\right)|\hat{\bS}|^{\frac{\nu_0+T}{2}}},
\end{align*}
where the shrinkage hyperparameter $\kappa$ appears in $\bV_{\bA}$, and the last equality holds because 
\begin{align*}
	\int |\vSigma|^{-\frac{\nu_0+T+n+k+1}{2}} & 
	\e^{-\frac{1}{2}\text{tr}\left(\vSigma^{-1}\hat{\bS}\right)}\e^{-\frac{1}{2}\text{tr}\left(\vSigma^{-1}(\bA-\hat{\bA})'\bK_{\bA}^{-1}(\bA-\hat{\bA})\right)}\di(\bA,\vSigma) \\
	& = (2\pi)^{\frac{nk}{2}}2^{\frac{n(\nu_0+T)}{2}}|\bK_{\bA}^{-1}|^{\frac{n}{2}}
			\Gamma_n\left(\frac{\nu_0+T}{2}\right)|\hat{\bS}|^{-\frac{\nu_0+T}{2}}.
\end{align*}
Finally, the marginal density $p(\bh \gvn \phi)$ has the following analytical expression:
\begin{align*}
	p(\bh\gvn \phi) & = \int p(\bh \gvn \phi,\sigma^2)p(\sigma^2)\di\sigma^2 \\
	& = (2\pi)^{-\frac{T}{2}}|1-\phi^2|^{\frac{1}{2}}S_{\sigma^2}^{\nu_{\sigma^2}}\Gamma(\nu_{\sigma^2})^{-1} 
	\int (\sigma^{2})^{-(\nu_{\sigma^2} + \frac{T}{2}+1)}\e^{-\frac{1}{\sigma^2} \widetilde{S}_{\sigma^2}}\di\sigma^2 \\
	& = (2\pi)^{-\frac{T}{2}}|1-\phi^2|^{\frac{1}{2}}\frac{\Gamma\left(\nu_{\sigma^2}+\frac{T}{2}\right)
	S_{\sigma^2}^{\nu_{\sigma^2}}}{\Gamma(\nu_{\sigma^2})\widetilde{S}_{\sigma^2}^{\nu_{\sigma^2}+\frac{T}{2}}},
\end{align*}
where $\Gamma(\cdot)$ is the gamma function and $ \widetilde{S}_{\sigma^2}$ is defined in Step 4 of the MCMC algorithm.

\subsection*{Cholesky Stochastic Volatility}

Next, we outline the estimation of the VAR-SV model:
\begin{equation} \label{Appendix_eq:varsv}
	\by_t = \ba_0 + \bA_1 \by_{t-1} + \cdots + \bA_p\by_{t-p} + \vepsilon_t, 	\quad \vepsilon_t \sim\distn{N}(\mathbf{0}, \vSigma_t),		
\end{equation}
where $\vSigma_t^{-1} = \bB_0' \bD_t^{-1}\bB_0$, $\bD_t = \diag(\e^{h_{1,t}}, \ldots, \e^{h_{n,t}})$ and $\bB_{0}$ is an $n \times n$ lower triangular matrix with ones on the diagonal. Each element of $\bh_t = (h_{1,t}, \ldots, h_{n,t})'$ follows an autoregressive process:
\[
	h_{i,t} = \mu_i + \phi_i(h_{i,t-1}-\mu_i) + u_{i,t}^h, \quad u_{i,t}^h \sim \distn{N}(0, \sigma_{i}^2)
\]
for $t=2,\ldots, T$, and the initial condition is specified as $h_{i,1}\sim\distn{N}(\mu_i,\sigma^2/(1-\phi_i^2))$. 

We aim to estimate the system equation by equation, which substantially speeds up the computation time. Recall that $\valpha_i$ denotes the $k\times 1$ vector that consists of the intercept and VAR coefficients in the $i$-th equation, and $\vbeta_i $ represents the $(i-1)\times 1$ vector of free elements in the $i$-th row of the impact matrix $\bB_0$. Then, the parameters for the $i$-th equation are $\valpha_i, \vbeta_i, \mu_i, \phi_i,$ and $\sigma_i^2$. We assume the following independent prior distributions on the parameters:
\begin{equation}\label{eq:VARSV-prior}
\begin{split}
	(\valpha_i\gvn \vkappa) & \sim\distn{N}(\valpha_{0,i},\bV_{\valpha_i}), \; 
	(\vbeta_i \gvn \vkappa) \sim\distn{N}(\vbeta_{0,i},\bV_{\vbeta_i}), \\
	\mu_i & \sim \distn{N}(\mu_{0,i},V_{\mu_i}), \; \phi_i\sim \distn{N}(\phi_{0,i},V_{\phi_i})1(|\phi_i|<1),\; 	\sigma_{i}^2 \sim \distn{IG}(\nu_{i},S_{i}),
\end{split}
\end{equation}
where $ \vkappa$ is a vector of hyperparameters that is described in more detail below. The prior mean vector $\valpha_{0,i}$ and the prior covariance matrix $\bV_{\valpha_i}$ are selected to mimic the Minnesota prior. More specifically, we set $\valpha_{0,i} = \mathbf{0}$ to shrink the VAR coefficients to zero. For $\bV_{\valpha_i}$, we assume it to be diagonal with the $k$-th diagonal element $V_{\valpha_i, kk}$ set to be:
\[
	V_{\valpha_i,kk} = \left\{
	\begin{array}{ll}
			\frac{\kappa_1}{l^2}, & \text{for the coefficient on the $l$-th lag of variable } i,\\
			\frac{\kappa_2 s_i^2}{l^2 s_j^2}, & \text{for the coefficient on the $l$-th lag of variable } j, j\neq i, \\
			100 s_i^2, & \text{for the intercept}, \\
	\end{array} \right.
\]
where $s_r^2$ denotes the sample variance of the residuals from an AR(4) model for the variable~$r$ for $r=1,\ldots, n$. In addition, we set the prior mean vector $\vbeta_{0,i}$ to be zero to shrink the impact matrix to the identity matrix. In addition, the prior covariance matrix $\bV_{\vbeta_i}$ is assumed to be diagonal, where the $j$-th diagonal element is set to be $\kappa_3 s_i^2/s_j^2$. Finally, we treat the shrinkage hyperparameters $\vkappa = (\kappa_1,\kappa_2,\kappa_3)'$ as unknown parameters to be estimated with hierarchical gamma priors $\kappa_i\sim\distn{G}(c_{j,1},c_{j,2}), j=1,2,3.$

Let $\by_{i,\bigcdot} = (y_{i,1},\ldots, y_{i,T})'$ denote the vector of observed values for the $i$-th variable for $i=1,\ldots, n$. We similarly define $\bh_{i,\bigcdot} = (h_{i,1},\ldots, h_{i,T})'$
. Next, stack $\by = (\by_{1,\bigcdot}',\ldots, \by_{n,\bigcdot}')', \bh = (\bh_{1,\bigcdot}',\ldots, \bh_{n,\bigcdot}')'$ and $\valpha = (\valpha_1',\ldots, \valpha_n')'$; similarly define $\vbeta, \vmu,\vphi$ and $\vsigma^2$. 
Then, posterior draws can be obtained by sampling sequentially from: 
\begin{enumerate}
	\item $p(\valpha \gvn \by,\vbeta, \bh, \vmu, \vphi,\vsigma^2,\vkappa)$; 
	\item $p(\vbeta \gvn \by,\valpha, \bh, \vmu, \vphi,\vsigma^2,\vkappa) = \prod_{i=2}^n p(\vbeta_i \gvn \by, \valpha, \bh_{i,\bigcdot}, \vkappa) $; 
		
	\item $p(\bh \gvn \by, \valpha,\vbeta, \vmu, \vphi,\vsigma^2,\vkappa) 
	= \prod_{i=1}^n p(\bh_{i,\bigcdot} \gvn \by, \valpha,\vbeta, \mu_i,\phi_i,\sigma^2_i)$; 
	\item $p(\vmu \gvn \by, \valpha,\vbeta, \bh, \vphi, \vsigma^2,\vkappa) = \prod_{i=1}^n p(\mu_i \gvn \bh_{i,\bigcdot}, \phi_i, \sigma^2_i) $; 
	\item $p(\vphi \gvn \by, \valpha,\vbeta, \bh, \vmu, \vsigma^2,\vkappa) = \prod_{i=1}^n p(\phi_i \gvn \bh_{i,\bigcdot}, \mu_i, \sigma^2_i) $; 
	\item $p(\vsigma^2 \gvn \by, \valpha,\vbeta, \bh, \vmu, \vphi,\vkappa) = \prod_{i=1}^n p(\sigma_i^2 \gvn \bh_{i,\bigcdot}, \mu_i, \phi_i)$;
	
	\item $p(\vkappa\gvn \by, \valpha, \vbeta, \bh, \vmu, \vphi,\vsigma^2) = p(\vkappa\gvn \valpha,\vbeta). $
\end{enumerate}

\textbf{Step 1}: we sample the reduced-form VAR coefficients equation by equation using the triangular algorithm proposed in \citet{CCCM22}. To that end, define $\bA_{i=0}$ to be a $k\times n$ matrix that has exactly the same elements as $\bA = (\ba_0, \bA_1,\ldots, \bA_p)'$ except for the $i$-th column, which is set to be zero, i.e., $\bA_{i=0} = (\valpha_1,\ldots, \valpha_{i-1}, \mathbf{0}, \valpha_{i+1},\ldots, \valpha_n).$ Then, we can rewrite~\eqref{Appendix_eq:varsv} as
\[
	\bB_0(\by_t - \bA_{i=0}'\bx_t) = \bB_{0,1:n,i}\valpha_i' \bx_t + \tilde{\vepsilon}_t, \quad 
	\tilde{\vepsilon}_t \equiv \bB_0\vepsilon_t\sim\distn{N}(\mathbf{0},\bD_t),
\]
where $\bx_t = (1,\by_{t-1}',\ldots, \by_{t-p}')'$ and $\bB_{0,1:n,i}$ is the $i$-th column of $\bB_0$. Transposing and stacking the above equation over $t=1,\ldots, T,$, we can write the system more compactly as
\[
	(\bY - \bX\bA_{i=0})\bB_0' = \bX\valpha_i\bB_{0,1:n,i}' +  \tilde{\vepsilon},
\]
where the matrices $\bY$, $\bX$ and $\tilde{\vepsilon}$ are respectively of dimensions $T\times n$, $T\times k$ and $T\times n$. Vectoring the above system, we obtain
\[
	\bz^i = \bW^i \valpha_i + \text{vec}(\tilde{\vepsilon}), \quad  \text{vec}(\tilde{\vepsilon})\sim\distn{N}(\mathbf{0},\bD),
\]
where $\bz^i_t = \text{vec}((\bY - \bX\bA_{i=0})\bB_0')$, $\bW^i = \bB_{0,1:n,i}\otimes\bX$ and $\bD = \text{diag}(\bD_{1,\bigcdot}, \ldots, \bD_{n,\bigcdot} )$ with $\bD_{i,\bigcdot} = \text{diag}(\e^{h_{i,1}},\ldots, \e^{h_{i,T}}).$ Given the conditionally Gaussian prior on $\valpha_i $ in~\eqref{eq:VARSV-prior} (conditional on $\vkappa$), it follows from standard linear regression results that
\[
	(\valpha_i \gvn \by, \vbeta, \valpha_{-i}, \bh, \vmu, \vphi, \vsigma^2, \vkappa) 
	\sim \distn{N}(\hat{\valpha}_i,\bK_{\valpha_i}^{-1}),
\]
where $\valpha_{-i}= (\valpha_{1}',\ldots, \valpha_{i-1}', \valpha_{i+1}',\ldots, \valpha_{n}')'$,
\[
	\bK_{\valpha_i} = \bV_{\valpha_i}^{-1} + \bW^{i'}\bD^{-1}\bW^i, \quad 
	\hat{\valpha}_i = \bK_{\valpha_i}^{-1}\left(\bV_{\valpha_i}^{-1}\valpha_{0,i} + \bW^{i'}\bD^{-1}\bz^i\right).
\]
Some of the above computations can be simplified by noting that $\bB_0$ is a lower triangular matrix. In addition, random draws from the high-dimensional $\distn{N}(\hat{\valpha}_i,\bK_{\valpha_i}^{-1})$ distribution can be obtained efficiently without inverting any large matrices; see, e.g., \citet{chan21} for computational details. 

\textbf{Step 2}: to sample $\vbeta$, the lower triangular elements of $\bB_0$, first note that given $\by$ and $\valpha$, the innovations $\vepsilon_1, \ldots, \vepsilon_T$ can be computed using~\eqref{Appendix_eq:varsv}. Next, we rewrite the model as a system of $(n-1)$ regressions in which $\epsilon_{i,t}$ is regressed
on the negative values of $\epsilon_{1,t},\ldots, \epsilon_{i-1,t}$ for $i = 2, \ldots , n$, and
$B_{0,i,1}, \ldots, B_{0,i,i-1} $ are the corresponding regression coefficients. More specifically, note that
\begin{align*}
    \bB_0 \vepsilon_{t} & =
    \begin{pmatrix}
        1   & 0 & 0 & \cdots    & 0 \\
        B_{0,2,1}  & 1 & 0 & \cdots    & 0 \\
        B_{0,3,1}  & B_{0,3,2} & 1 & \cdots    & \vdots \\
        \vdots  & \vdots&   & \ddots    & \vdots \\
        B_{0,n,1}  & B_{0,n,2} & B_{0,n,3} & \cdots    & 1
    \end{pmatrix}
    \begin{pmatrix}
            \epsilon_{1,t}\\
            \epsilon_{2,t}\\
            \epsilon_{3,t} \\
            \vdots \\
            \epsilon_{n,t}
    \end{pmatrix} =
    \begin{pmatrix}
            \epsilon_{1,t}\\
            \epsilon_{2,t} + B_{0,2,1}\epsilon_{1,t}\\
            \epsilon_{3,t} + B_{0,3,1}\epsilon_{1,t} + B_{0,3,2}\epsilon_{2,t} \\
            \vdots \\
            \epsilon_{n,t} + \sum_{j=1}^{n-1}B_{0,n,j}\epsilon_{j,t}
    \end{pmatrix} \\
     & = \begin{pmatrix}
            \epsilon_{1,t}\\
            \epsilon_{2,t}\\
            \epsilon_{3,t} \\
            \vdots \\
            \epsilon_{n,t}
    \end{pmatrix} -
\begin{pmatrix}
0 & 0 & 0 &  0 & 0 & \cdots & \cdots & 0 \\
- \epsilon_{1,t} & 0 & 0 & 0 & 0 & \cdots &  & \vdots \\
0 & -\epsilon_{1,t} & -\epsilon_{2,t} & 0 & 0 & \cdots & & 0 \\
\vdots &  & & \ddots & \ddots & & \cdots & 0 \\
0 & \cdots & 0 & \cdots & 0 & -\epsilon_{1,t} & \cdots & - \epsilon_{t,n-1}
\end{pmatrix}
\begin{pmatrix}
 B_{0,2,1}\\
 B_{0,3,1} \\
 B_{0,3,2} \\
 \vdots \\
 B_{0,n,n-1}
\end{pmatrix}.
\end{align*}
Or more succinctly,
\[
	 \bB_0 \vepsilon_{t} = \vepsilon_t - \bE_t\vbeta.
\]
Since $|\vSigma_t| = |\bD_t|$, we can rewrite the likelihood implied by \eqref{Appendix_eq:varsv} as
\begin{align*}
    p(\by \gvn \valpha, \vbeta, \bh) &  \propto \left(\prod_{t=1}^{T}|\bD_t|^{-\frac{1}{2}}\right)
    \exp \left(-\frac{1}{2}\sum_{t=1}^{T}\vepsilon_t'(\bB_0'\bD_t^{-1}\bB_0)\vepsilon_t \right) \\
    & = \left(\prod_{t=1}^{T}|\bD_t|^{-\frac{1}{2}}\right)
    \exp \left(-\frac{1}{2}\sum_{t=1}^{T}(\vepsilon_t - \bE_t\vbeta)'\bD_t^{-1}( \vepsilon_t - \bE_t\vbeta) \right).
\end{align*}
That is, the likelihood is the same as that implied by the regression
\begin{equation} \label{eq:varsv-e}
	\vepsilon_t = \bE_t\vbeta + \vect{\eta}_t,  
\end{equation}
where $\vect{\eta}_t\sim\distn{N}(\mathbf{0}, \bD_t).$ In addition, since the covariance matrix $\bD_t$ is diagonal and $\bE_t$ is block-diagonal, we can sample $\vbeta_2,\ldots, \vbeta_n$ equation by equation to improve computational efficiency. 

To that end, we stack the $i$-th equation, $i=2,\ldots, n$, over $t=1,\ldots, T$ and obtain
\[
	\vepsilon_{i,\bigcdot} = \bE_{i,\bigcdot} \vbeta_i + \vect{\eta}_{i,\bigcdot},
\]
where $\vepsilon_{i,\bigcdot} = (\epsilon_{i,1},\ldots, \epsilon_{i,T})'$, 
$\bE_{i,\bigcdot} = (\vepsilon_{1,\bigcdot},\ldots, \vepsilon_{i-1,\bigcdot})$ and 
$\vect{\eta}_{i,\bigcdot}\sim\distn{N}(\mathbf{0}, \bD_{i,\bigcdot})$ with $\bD_{i,\bigcdot} = \diag(\e^{h_{i,1}},\ldots,\e^{h_{i,T}})$. Finally, given the hierarchical Gaussian prior $(\vbeta_i\gvn\vkappa) \sim\distn{N}(\vbeta_{0,i},\bV_{\vbeta_i})$, it follows that
\[
	(\vbeta_i \gvn \by, \valpha, \bh_{i,\bigcdot}, \vkappa)\sim\distn{N}(\hat{\vbeta}_i, \bK_{\vbeta_i}^{-1}),
\]
where
\[
	\bK_{\vbeta_i} = \bV_{\vbeta_i}^{-1} + \bE_{i,\bigcdot}'\bD_{i,\bigcdot}^{-1}\bE_{i,\bigcdot}, \quad
	\hat{\vbeta}_i = \bK_{\vbeta_i}^{-1}\left(\bV_{\vbeta_i}^{-1}\vbeta_{0,i} + \bE_{i,\bigcdot}'\bD_{i,\bigcdot}^{-1}\vepsilon_{i,\bigcdot} \right).
\]

\textbf{Step 3}: to sample $\bh$, we first compute the ``orthogonalized" innovations: $\tilde{\vepsilon}_t = \bB_0(\by_t - \ba_0 - \bA_1\by_{t-1} - \cdots - \bA_p\by_{t-p})$ for $t=1,\ldots, T$. Then, we can sample each vector $\bh_{i,\bigcdot}$ separately for $i=1,\ldots, n$. More specifically, using the time series $\tilde{\epsilon}_{i,1},\ldots, \tilde{\epsilon}_{i,T}$ as raw data, we directly apply the auxiliary mixture sampler of \citet*{KSC98} in conjunction with the precision sampler of \citet{CJ09} to sample $(\bh_{i,\bigcdot} \gvn \by, \valpha,\vbeta, \mu_i,\phi_i,\sigma^2_i)$ for $i=1,\ldots, n$. For a textbook treatment, see, e.g., Chapter 19 in \citet{CKPT19}.

\textbf{Step 4}: this step can be done easily, as $\mu_1,\ldots, \mu_n$ are conditionally independent given $\bh$ and other parameters, and each follows a normal distribution:
\[
	(\mu_i \gvn \bh_{i,\bigcdot}, \phi_i, \sigma^2_i)\sim \distn{N}(\hat{\mu}_i, K_{\mu_i}^{-1}),
\]
where
\begin{align*}
	K_{\mu_i} & = V_{\mu_i}^{-1} + \frac{1}{\sigma_i^2}\left[ 1-\phi_i^2 + (T-1)(1-\phi_i)^2\right] \\
	\hat{\mu}_i & = K_{\mu_i}^{-1}\left[V_{\mu_i}^{-1}\mu_{0,i} + \frac{1}{\sigma_i^2}\left( (1-\phi_i^2)h_{i,1} + (1-\phi_i)\sum_{t=2}^T(h_{i,t}-\phi_ih_{i,t-1})\right)\right].
\end{align*}

\textbf{Step 5}: to sample $\vphi$, first note that 
\[
	p(\phi_i \gvn \bh_{i,\bigcdot},\mu_i,\sigma^2_i)\propto p(\phi_i)g(\phi_i)\e^{-\frac{1}{2\sigma^2_i}\sum_{t=2}^T(h_{i,t}-\mu_i-\phi_i(h_{i,t-1}-\mu_i))^2},
\]
where $ g(\phi_i) = (1-\phi_i^2)^{1/2}\e^{-\frac{1}{2\sigma_i^2}(1-\phi_i^2)(h_{i,1}-\mu_i)^2}$ and $p(\phi_i)$ is the truncated normal prior given in~\eqref{eq:VARSV-prior}. The conditional density $p(\phi_i \gvn \bh_{i,\bigcdot},\mu_i,\sigma^2_i)$ is nonstandard, but a draw from it can be obtained by using an independence-chain Metropolis-Hastings step with proposal distribution $\distn{N}(\hat{\phi}_i, K_{\phi_i}) 1(|\phi_i|<1)$, where 
\begin{align*}
	K_{\phi_i}   & = V_{\phi_i}^{-1} + \frac{1}{\sigma_i^2}\sum_{t=2}^{T}(h_{i,t-1}-\mu_i)^2\\
	\hat{\phi}_i & = K_{\phi_i}^{-1}\left[V_{\phi_i}^{-1}\phi_{0,i} + \frac{1}{\sigma_i^2}\sum_{t=2}^{T}(h_{i,t-1}-\mu_i) (h_{i,t}-\mu_i) \right].
\end{align*}
Then, given the current draw $\phi_i$, a proposal $\phi_i^*$ is accepted with probability $\min(1,g(\phi_i^*)/g(\phi_i))$; 
otherwise the Markov chain stays at the current state $\phi_i$.

\textbf{Step 6}: to sample $\sigma^2_1, \ldots, \sigma_n^2$, note that each follows an inverse-gamma distribution:
\[
	(\sigma_i^2 \gvn \bh_{i,\bigcdot},\mu_i,\phi_i) \sim \distn{IG}\left(\nu_{i}+\frac{T}{2}, \widetilde{S}_i\right),
\]
where $ \widetilde{S}_i = S_i + [(1-\phi_i^2)(h_{i,1}-\mu_i)^2 + \sum_{t=2}^T(h_{i,t}-\mu_i-\phi_i(h_{i,t-1}-\mu_i))^2]/2$.

\textbf{Step 7}: note that $\kappa_1, \kappa_2$ and $\kappa_3$ only appear in their priors $\kappa_j\sim \distn{G}(c_{j,1},c_{j,2}), j=1,2,3 $, and the prior covariance matrices of $\valpha_i$ and $\vbeta_i$ in \eqref{eq:VARSV-prior}.  To sample $\kappa_1, \kappa_2$ and $\kappa_3$, first define the index set $S_{\kappa_1}$ that collects all the indexes $(i,j)$ such that $\alpha_{i,j}$ is a coefficient associated with an own lag. That is, 
$S_{\kappa_1} = \{(i,j): \alpha_{i,j} \text{ is a coefficient associated with an own lag}\}$. Similarly, define $S_{\kappa_2}$ as the set that collects all the indexes $(i,j)$ such that $\alpha_{i,j}$ is a coefficient associated with a lag of other variables. Lastly, define $S_{\kappa_3} = \{(i,j): i=2,\ldots, n, j=1,\ldots, i-1\}$. It is easy to check that the numbers of elements in $S_{\kappa_1}$, $S_{\kappa_2}$ and $S_{\kappa_3}$ are, respectively, $np, (n-1)np$ and $n(n-1)/2$. Then, we have
\begin{align*}
	p(\kappa_{1} \gvn \valpha) & \propto \prod_{(i,j)\in S_{\kappa_1}} \kappa_1^{-\frac{1}{2}} \e^{-\frac{1}{2\kappa_{1}C_{i,j}}(\alpha_{i,j}-\alpha_{0,i,j})^2}	\times \kappa_1^{c_{1,1}-1}\e^{-\kappa_1 c_{1,2}} \\
	 & = \kappa_1^{c_{1,1}-\frac{np}{2}-1} \e^{-\frac{1}{2} \left(2c_{1,2}\kappa_1 + \kappa_{1}^{-1}\sum_{(i,j)\in S_{\kappa_1}}\frac{(\alpha_{i,j}-\alpha_{0,i,j})^2}{C_{i,j}}\right)},
\end{align*}
where $\alpha_{0,i,j}$ is the $j$-th element of the prior mean vector $\valpha_{0,i}$ and $C_{i,j}$ is a constant determined by the Minnesota prior. Note that the above expression is the kernel of the $\distn{GIG}\left(c_{1,1}-\frac{np}{2}, 2c_{1,2}, \sum_{(i,j)\in S_{\kappa_1}}\frac{(\alpha_{i,j}-\alpha_{0,i,j})^2}{C_{i,j}}\right)$ distribution. Similarly, we have
\begin{align*}
	(\kappa_2 \gvn \valpha) & \sim \distn{GIG}\left(c_{2,1}-\frac{(n-1)np}{2}, 2c_{2,2},\sum_{(i,j)\in S_{\kappa_2}} \frac{(\alpha_{i,j}-\alpha_{0,i,j})^2}{C_{i,j}}\right), \\
	(\kappa_3 \gvn \vbeta) & \sim \distn{GIG}\left(c_{3,1}-\frac{n(n-1)}{4}, 2c_{3,2},\sum_{(i,j)\in S_{\kappa_3}} \frac{(\beta_{i,j}-\beta_{0,i,j})^2}{\tilde{C}_{i,j}}\right),
\end{align*}
where $\tilde{C}_{i,j} = s_i^2/s_j^2$ and $s_r^2$ is the sample variance of the residuals from an AR(4) model for variable $r$.

Next, we provide the details of estimating the marginal likelihood of the VAR-SV model. As before, our marginal likelihood estimator of $p(\by)$ has two parts: the conditional Monte Carlo part in which we integrate out the VAR coefficients $\valpha$; and the adaptive importance sampling part that biases the joint distribution of $\vbeta, \bh, \vmu, \vphi, \vsigma^2$ and $\vkappa$. In what follows, we first derive an analytical expression of the conditional Monte Carlo estimator $\Em[p(\by \gvn\vtheta, \bh, \vkappa) \gvn \vbeta, \bh,\vkappa] = p(\by\gvn \vbeta, \bh, \vkappa)$. To that end, rewrite~\eqref{Appendix_eq:varsv} as
\[
	\bY\bB_0' = \bX\bA\bB_0' + \tilde{\vepsilon},
\]
where $\bA = (\ba_0,\bA_1, \ldots, \bA_p)'$, $ \tilde{\vepsilon}=\vepsilon\bB_0'$ and the matrices $\bY$, $\bX$ and $\vepsilon$ are, respectively, of dimensions $T\times n$, $T\times k$ and $T\times n$. Vectoring the above system, we obtain
\[
	\tilde{\by} = \tilde{\bX} \valpha + \text{vec}(\tilde{\vepsilon}), \quad  \text{vec}(\tilde{\vepsilon})\sim\distn{N}(\mathbf{0},\bD),
\]
where $\tilde{\by} = \text{vec}(\bY\bB_0'),$ $\tilde{\bX} = \bB_0\otimes\bX$ and $\valpha = \text{vec}(\bA) = (\valpha_{1}',\ldots, \valpha_{n}')'$. Furthermore, let $\valpha_0 = (\valpha_{0,1}',\ldots, \valpha_{0,n}')'$ and $\bV_{\valpha} =\text{diag}(\bV_{\valpha_1},\ldots,\bV_{\valpha_n})$ and $k_2 = (2\pi)^{-\frac{nT+nk}{2}}\e^{-\frac{1}{2}\mathbf{1}_{nT}'\bh}|\bV_{\valpha}|^{-\frac{1}{2}}$. Then, we have
\begin{align*}
	p(\by\gvn\vbeta, \bh, \vkappa) & = \int p(\by \gvn \vbeta, \valpha,\bh)p(\valpha\gvn \vkappa) \di\valpha \\
	& = k_2 \int\e^{-\frac{1}{2}(\tilde{\by}-\tilde{\bX}\valpha)'\bD^{-1}(\tilde{\by}-\tilde{\bX}\valpha)
	-\frac{1}{2}(\valpha-\valpha_{0})'\bV_{\valpha}^{-1}(\valpha-\valpha_{0})} \di\valpha  \\	
	&= k_2 \e^{-\frac{1}{2}\left(\tilde{\by}'\bD^{-1}\tilde{\by} + \valpha_{0}'\bV_{\valpha}^{-1}\valpha_{0} - \hat{\valpha}'\bK_{\valpha}\hat{\valpha} \right)} \int\e^{-\frac{1}{2}(\valpha-\hat{\valpha})'\bK_{\valpha}(\valpha-\hat{\alpha})}\di\valpha  \\
	& = (2\pi)^{-\frac{nT}{2}}\e^{-\frac{1}{2}\mathbf{1}_{nT}'\bh}	|\bV_{\valpha}|^{-\frac{1}{2}}|\bK_{\valpha}|^{-\frac{1}{2}}\e^{-\frac{1}{2}\left(\tilde{\by}'\bD^{-1}\tilde{\by} + \valpha_{0}'\bV_{\valpha}^{-1}\valpha_{0} 		- \hat{\valpha}'\bK_{\valpha}\hat{\valpha}\right)},
\end{align*}
where 
\[
	\bK_{\valpha} = \bV_{\valpha}^{-1} + \tilde{\bX}'\bD^{-1}\tilde{\bX}, \quad 
	\hat{\valpha} = \bK_{\valpha}^{-1}\left(\bV_{\valpha}^{-1}\valpha_0 + \tilde{\bX}'\bD^{-1}\tilde{\by}\right).
\]
The last equality holds because 
\[
	\int\e^{-\frac{1}{2}(\valpha-\hat{\valpha})'\bK_{\valpha}(\valpha-\hat{\valpha})}\di\valpha	= (2\pi)^{\frac{nk}{2}}|\bK_{\valpha}|^{-\frac{1}{2}}.
\]
Note that the vector of shrinkage hyperparameters $\vkappa$ appears in the prior covariance matrix $\bV_{
\valpha}$.\footnote{One potential limitation of this conditional Monte Carlo method is that it requires the computation of the precision matrix $\bK_{\valpha}$. For very large VARs $(n\approx 100)$, the computational and memory requirements for this step might become excessive. One feabile way forward is to approximate this precision matrix using some low-rank decompostion. This possibility is left to future research.} Now, we can write the marginal likelihood as:
\begin{align*}
	p(\by) & = \int p(\by \gvn \vbeta, \bh, \vkappa) p(\vkappa)\prod_{i=1}^n p(\vbeta_i)p(\bh_{i,\bigcdot}\gvn\mu_i,\phi_i,\sigma^2_i)p(\mu_i)p(\phi_i)p(\sigma^2_i)\di(\vbeta,\bh,\vmu,\vphi,\vsigma^2,\vkappa) \\
	&=  \int p(\by\gvn\vbeta, \bh,\vkappa) p(\vkappa)\prod_{i=1}^n p(\vbeta_i) p(\bh_{i,\bigcdot}\gvn\mu_i,\phi_i)p(\mu_i)p(\phi_i)\di(\vbeta,\bh,\vmu,\vphi,\vkappa), 
\end{align*}
where $p(\bh_{i,\bigcdot}\gvn\mu_i,\phi_i,\sigma^2_i)$ is a Gaussian density implied by the state equation \eqref{eq:hit}, $p(\mu_i), p(\phi_i)$, $p(\sigma^2_i)$ and $p(\vkappa)$ are the prior densities. Furthermore, the density $p(\bh_{i,\bigcdot}\gvn\mu_i,\phi_i)$ has the following analytical expression:
\begin{align}
	p(\bh_{i,\bigcdot}\gvn\mu_i,\phi_i) & = \int p(\bh_{i,\bigcdot}\gvn\mu_i,\phi_i,\sigma^2_i)p(\sigma^2_i)\di\sigma^2_i  \nonumber \\
	& = (2\pi)^{-\frac{T}{2}}|1-\phi_i^2|^{\frac{1}{2}}S_i^{\nu_i}\Gamma(\nu_i)^{-1} \int (\sigma^{2}_i)^{-(\nu_i + \frac{T}{2}+1)}\e^{-\frac{ \widetilde{S}_i}{\sigma_i^2}}\di\sigma^2_i,\nonumber	\\
	& = (2\pi)^{-\frac{T}{2}}|1-\phi_i^2|^{\frac{1}{2}}\frac{\Gamma\left(\nu_i+\frac{T}{2}\right)S_i^{\nu_i}}{\Gamma(\nu_i)\widetilde{S}_i^{\nu_i+\frac{T}{2}}}, \label{eq:den_hmarg}
\end{align}
where $\Gamma(\cdot)$ is the gamma function and $ \widetilde{S}_i = S_i + [(1-\phi_i^2)(h_{i,1}-\mu_i)^2 + \sum_{t=2}^T(h_{i,t}-\mu_i-\phi_i(h_{i,t-1}-\mu_i))^2]/2$.

Next, we combine the conditional Monte Carlo with the importance sampling approach described in Section~\ref{ss:IS}. Specifically, we consider the parametric family
\[
	\mathcal{F} = \left\{
	\prod_{i=1}^n f(\vbeta_i; \hat{\vbeta}_i,\hat{\bK}_{\vbeta_{i}}^{-1}) f_{\distn{N}}(\bh_{i,\bigcdot}; \hat{\bh}_{i,\bigcdot}, \hat{\bK}_{\bh_{i,\bigcdot}}^{-1})	f_{\distn{N}}(\mu_i; \hat{\mu}_i, \hat{K}_{\mu_i}^{-1})	f_{\distn{N}}(\phi_i; \hat{\phi}_i, \hat{K}_{\phi_i}^{-1})	\prod_{j=1}^3 f_{\distn{G}}(\kappa_j; \hat{\nu}_{\kappa_j}, \hat{S}_{\kappa_j})	
	\right\}.
\]
The parameters of the importance sampling densities are chosen by solving the maximization problem in~\eqref{eq:maxMC}. In particular, the $T$-variate Gaussian density is obtained using the method described in Section~\ref{ss:CSV}. The other importance sampling densities can be obtained similarly.
 
\subsection*{Factor Stochastic Volatility}

In this section we describe the posterior sampler for estimating the VAR-FSV model:
\begin{align*}
	\by_t       & = \ba_0 + \bA_1 \by_{t-1} + \cdots + \bA_p\by_{t-p} + \vepsilon_t, \\
	\vepsilon_t & = \bL \mathbf{f}_t + \bu_t^y, \quad \bu_t^y\sim\distn{N}(\mathbf{0},\bD_t),
\end{align*}
where $\mathbf{f}_t = (f_{1,t},\ldots, f_{r,t})'$ is a $r\times 1$ vector of latent factors at time $t$, $\bL$ is the associated $n\times r$ matrix of factor loadings and $\bD_t = \text{diag}(\e^{h_{1,t}},\ldots, \e^{h_{n,t}})$. The factors are serially independent and distributed as $\mathbf{f}_t \sim \distn{N}(\mathbf{0},\bG_t)$, where $\bG_t = \text{diag}(\e^{h_{n+1,t}},\ldots, \e^{h_{n+r,t}})$ is diagonal. For identification purposes we assume $\bL$ is a lower triangular matrix with ones on the main diagonal. Finally, for $i=1,\ldots, n+r$, the log-volatility evolves as:
\[
	h_{i,t} = \mu_{i} + \phi_i(h_{i,t-1} - \mu_i) + u_{i,t}^h, \quad  u_{i,t}^h\sim\distn{N}(0,\sigma_i^2)
\]
for $t=2,\ldots, T$, and the initial conditions are assumed to follow the stationary distribution $h_{i,1} \sim \distn{N}(\mu_i,\sigma_i^2/(1-\phi_i^2))$. 

Next, we specify the prior distributions on the model parameters. Let $\valpha_i$ and $\bl_i$ denote the VAR coefficients and the free elements of $\bL$ in the $i$-th equation, respectively, for $i=1,\ldots, n$. We assume the following independent priors on $\valpha_i$ and $\bl_i$ for $i=1,\ldots, n$:
\[
	\valpha_i\sim\distn{N}(\valpha_{0,i},\bV_{\valpha_i}),\quad \bl_i\sim\distn{N}(\bl_{0,i},\bV_{\bl_i}).
\]
We set the prior mean $\bl_{0,i}$ to be zero and the prior covariance matrix $\bV_{\bl_i}$ to be the identity matrix. As before, the prior mean and prior covariance matrix are elicited according to the Minnesota prior. In particular, we set $\valpha_{0,i}=\mathbf{0}$ and $\bV_{\valpha_i}$ is set to be diagonal with the $k$-th diagonal element $V_{\valpha_i, kk}$:
\[
	V_{\valpha_i,kk} = \left\{
	\begin{array}{ll}
			\frac{\kappa_1}{l^2}, & \text{for the coefficient on the $l$-th lag of variable } i,\\
			\frac{\kappa_2 s_i^2}{l^2 s_j^2}, & \text{for the coefficient on the $l$-th lag of variable } j, j\neq i, \\			
			100 s_i^2, & \text{for the intercept}, \\
	\end{array} \right.
\]
where $s_r^2$ denotes the sample variance of the residuals from an AR(4) model for the variable~$r, r=1,\ldots, n$. Again the shrinkage hyperparameters $\vkappa = (\kappa_1,\kappa_2)'$ are treated as unknown parameters to be estimated with hierarchical gamma priors $\kappa_j\sim\distn{G}(c_{j,1},c_{j,2}), j=1,2.$ For the parameters in the stochastic volatility equations, we assume the same priors as in~\eqref{eq:VARSV-prior}:
\[
	\mu_i \sim \distn{N}(\mu_{0,i},V_{\mu_i}), \; \phi_i\sim \distn{N}(\phi_{0,i},V_{\phi_i})1(|\phi_i|<1),\; 	
	\sigma_{i}^2 \sim \distn{IG}(\nu_{i},S_{i}), \; i=1,\ldots, n.
\]

For notational convenience, stack $\by = (\by_1',\ldots, \by_T')', \mathbf{f} = (\mathbf{f}_1',\ldots, \mathbf{f}_T')', \bh = (\bh_1',\ldots, \bh_{n+r}')'$,  $\valpha = (\valpha_1',\ldots, \valpha_n')'$ and $\bl = (\bl_1',\ldots,\bl_n')'$; similarly define $\vmu,\vphi$ and $\vsigma^2$. In addition, let $\by_{i,\bigcdot} = (y_{i,1},\ldots, y_{i,T})'$ denote the vector of observed values for the $i$-th variable for $i=1,\ldots, n$. We similarly define $\bh_{i,\bigcdot} = (h_{i,1},\ldots, h_{i,T})', 
i=1,\ldots, n+r$. Then, posterior draws can be obtained by sampling sequentially from: 
\begin{enumerate}
	\item $p(\mathbf{f} \gvn \by, \bl,\valpha, \bh, \vmu, \vphi,\vsigma^2,\vkappa) 
	= p(\mathbf{f} \gvn \by, \bl, \valpha, \vrho, \bh)$; 
	
	\item $p(\valpha,\bl \gvn \by, \mathbf{f}, \bh, \vmu, \vphi,\vsigma^2, \vkappa) = 
	\prod_{i=1}^n p(\valpha_i,\bl_i \gvn \by_{i,\bigcdot}, \mathbf{f}, \bh_{i,\bigcdot}, \vkappa)$; 
	
	\item $p(\bh \gvn \by, \mathbf{f},\valpha,\bl, \vmu, \vphi,\vsigma^2, \vkappa) = 
	\prod_{i=1}^{n+r} p(\bh_{i,\bigcdot} \gvn \by, \mathbf{f},\valpha, \bl, \vmu, \vphi,\vsigma^2)$; 
	
	\item $p(\vsigma^2 \gvn \by, \mathbf{f}, \valpha, \bl, \bh, \vmu, \vphi, \vkappa) 	
	= \prod_{i=1}^{n+r} p(\sigma_i^2 \gvn \bh_{i,\bigcdot}, \mu_i, \phi_i)$;
	
	\item $p(\vmu \gvn \by, \mathbf{f}, \valpha, \bl, \bh, \vphi,\vsigma^2, \vkappa) 
	= \prod_{i=1}^{n+r} p(\mu_i \gvn \bh_{i,\bigcdot}, \phi_i, \sigma^2_i) $; 
	
	\item $p(\vphi \gvn \by, \mathbf{f}, \valpha, \bl, \bh, \vmu,\vsigma^2, \vkappa) 
	= \prod_{i=1}^{n+r} p(\phi_i \gvn \bh_{i,\bigcdot}, \mu_i, \sigma^2_i) $;
	
	\item $p(\vkappa \gvn \by, \mathbf{f}, \bl,\valpha, \bh, \vmu, \vphi,\vsigma^2) 
	= p(\vkappa \gvn \valpha)$.	
	
\end{enumerate}

\textbf{Step 1}: to sample $\mathbf{f}$, we first stack the VAR over $t=1,\ldots, T$ and write it as:
\[
	\vepsilon = (\mathbf{I}_T\otimes \bL)\mathbf{f} +  \bu^y, \quad 
	\bu^y \sim \distn{N}(\mathbf{0},\bD),
\]	
where $\vepsilon = (\vepsilon_1',\ldots, \vepsilon_T')'$ is known given $\valpha$ and the data
and $\bD = \text{diag}(\bD_1,\ldots, \bD_T)$. In addition, the prior on the factors can be written as $\mathbf{f} \sim \distn{N}(\mathbf{0},\bG)$, where $\bG = \text{diag}(\bG_1,\ldots, \bG_T)$. It follows from standard linear regression results that \citep[see, e.g.,][chapter 12]{CKPT19}
\[
	(\mathbf{f} \gvn\by, \valpha, \bl, \bh) \sim  \distn{N}(\hat{\mathbf{f}},\bK_{\mathbf{f}}^{-1}), 
\]
where
\begin{align*}
	\bK_{\mathbf{f}} = \bG^{-1} + (\mathbf{I}_T\otimes \bL')\bD^{-1}(\mathbf{I}_T\otimes \bL), \quad 
	\hat{\mathbf{f}}  = \bK_{\mathbf{f}}^{-1}(\mathbf{I}_T\otimes \bL')\bD^{-1}\vepsilon.
\end{align*}
Note that the precision matrix $\bK_{\mathbf{f}}$ is banded, and we can use the precision sampler of 
\citet{CJ09} to sample $\mathbf{f}$ efficiently.

\textbf{Step 2}: to sample $\valpha$ and $\bl$ jointly, first note that given the latent factors $\mathbf{f}$, the VAR becomes $n$ unrelated regressions and we can sample $\valpha$ and $\bl$ equation by equation. Recall that $\by_{i,\bigcdot} = (y_{i,1},\ldots, y_{i,T})'$ is defined to be the $T\times 1$ vector of observations for the $i$-th variable; $\valpha_i$ and $\bl_i$ denote, respectively, the VAR coefficients and the free element of $\bL$ in the $i$-th equation. Note that the dimension of $\bl_i$ is $i-1$ for $i\leq r$ and $r$ for $i > r$. Then, the $i$-th equation of the VAR can be written as
\begin{align*}
	\by_{i,\bigcdot} & = \bX_i\valpha_i + \mathbf{F}_{1:i-1}\bl_i + \mathbf{f}_{i,\bigcdot} + \bu_{i,\bigcdot}^y, \quad i\leq r, \\
	\by_{i,\bigcdot} & = \bX_i\valpha_i + \mathbf{F}_{1:r}\bl_i + \bu_{i,\bigcdot}^y, \quad i> r,
\end{align*}
where $\mathbf{f}_{i,\bigcdot} = (f_{i,1},\ldots, f_{i,T})'$ is the $T\times 1$ vector of the $i$-th factor and $\mathbf{F}_{1:j} = (\mathbf{f}_{1,\bigcdot},\ldots, \mathbf{f}_{j,\bigcdot})$
is the $T\times j$ matrix that contains the first $j$ factors. The vector of innovations $\bu_{i,\bigcdot}^y= (u_{i,1},\ldots, u_{i,T})'$ is distributed as $\distn{N}(\mathbf{0},\bD_{\bh_{i,\bigcdot}})$, where $\bD_{\bh_{i,\bigcdot}} =\text{diag}(\e^{h_{i,1}},\ldots, \e^{h_{i,T}})$. 
Letting $\vtheta_i = (\valpha_i',\bl_i')'$, we can further write the VAR systems as
\begin{equation}\label{eq:VAR-FSV}
	\tilde{\by}_{i,\bigcdot} =\bZ_i\vtheta_i + \bu_{i,\bigcdot}^y,
\end{equation}
where $\tilde{\by}_{i,\bigcdot} = \by_{i,\bigcdot} - \mathbf{f}_{i,\bigcdot} $ and $\bZ_i = (\bX_i,\mathbf{F}_{1:i-1})$ for $i\leq r$; $\tilde{\by}_i = \by_{i,\bigcdot}$ and $\bZ_i = (\bX_i,\mathbf{F}_{1:r})$ for $i > r$. Again using standard linear regression results, we obtain:
\[
	(\vtheta_i \gvn \by_{i,\bigcdot}, \mathbf{f}, \bh_{i,\bigcdot},\vkappa) \sim  \distn{N}(\hat{\vtheta}_i,\bK_{\vtheta_i}^{-1}), 
\]
where
\[
	\bK_{\vtheta_i} = \bV_{\vtheta_i}^{-1} + \bZ_i'\bD_{\bh_{i,\bigcdot}}^{-1}\bZ_i, \quad 
	\hat{\vtheta}_i  = \bK_{\vtheta_i}^{-1}(\bV_{\vtheta_i}^{-1}\vtheta_{0,i} +  \bZ_i\bD_{\bh_{i,\bigcdot}}^{-1}\tilde{\by}_{i,\bigcdot}) 
\]
with $\bV_{\vtheta_i} = \text{diag}(\bV_{\valpha_i},\bV_{\bl_i})$ and $\vtheta_{0,i} = (\valpha_{0,i}',\bl_{0,i}')'$.

\textbf{Step 3}: to sample $\bh$, again note that given the latent factors $\mathbf{f}$, the VAR becomes $n$ unrelated regressions and we can sample each vector $\bh_{i,\bigcdot} = (h_{i,1},\ldots, h_{i,T})'$ separately. More specifically, we can directly apply the auxiliary mixture sampler in \citet*{KSC98}  in conjunction with the precision sampler of \citet{CJ09} to sample from $(\bh_{i,\bigcdot} \gvn \by, \mathbf{f},\valpha,\bl,\vmu,\vphi,\vsigma^2)$ for $i=1,\ldots, n+r$. For a textbook treatment, see, e.g., \citet{CKPT19} chapter 19.

\textbf{Step 4}: this step can be done easily, as the elements of $\vsigma^2$ are conditionally independent and each follows an inverse-gamma distribution:
\[
	(\sigma_i^2 \gvn \bh_{i,\bigcdot},\mu_i,\phi_i) \sim \distn{IG}(\nu_{i}+T/2, \widetilde{S}_i),
\]
where $ \widetilde{S}_i = S_i + [(1-\phi_i^2)(h_{i,1}-\mu_i)^2 + \sum_{t=2}^T(h_{i,t}-\mu_i-\phi_i(h_{i,t-1}-\mu_i))^2]/2$.

\textbf{Step 5}: it is also straightforward to sample $\vmu$, as the elements of $\vmu$ are conditionally independent and each follows a normal distribution:
\[
	(\mu_i \gvn \bh_{i,\bigcdot}, \phi_i, \sigma^2_i)\sim \distn{N}(\hat{\mu}_i, K_{\mu_i}^{-1}),
\]
where
\begin{align*}
	K_{\mu_i} & = V_{\mu_i}^{-1} + \frac{1}{\sigma_i^2}\left[ 1-\phi_i^2 + (T-1)(1-\phi_i)^2\right] \\
	\hat{\mu}_i & = K_{\mu_i}^{-1}\left[V_{\mu_i}^{-1}\mu_{0,i} + \frac{1}{\sigma_i^2}\left( (1-\phi_i^2)h_{i,1} + (1-\phi_i)\sum_{t=2}^T(h_{i,t}-\phi_ih_{i,t-1})\right)\right].
\end{align*}

\textbf{Step 6}: to sample $\phi_i$ for $i=1,\ldots, n+r$, note that 
\[
	p(\phi_i \gvn \bh_{i,\bigcdot},\mu_i,\sigma^2_i)\propto p(\phi_i)g(\phi_i)\e^{-\frac{1}{2\sigma^2_i}\sum_{t=2}^T(h_{i,t}-\mu_i-\phi_i(h_{i,t-1}-\mu_i))^2},
\]
where $ g(\phi_i) = (1-\phi_i^2)^{\frac{1}{2}}\e^{-\frac{1}{2\sigma_i^2}(1-\phi_i^2)(h_{i,1}-\mu_i)^2}$ and $p(\phi_i)$ is the truncated normal prior. The conditional density $p(\phi_i \gvn \bh_{i,\bigcdot},\mu_i,\sigma^2_i)$ is nonstandard, but a draw from it can be obtained by using 
an independence-chain Metropolis-Hastings step with proposal distribution $\distn{N}(\hat{\phi}_i, K_{\phi_i}^{-1}) 1(|\phi_i|<1)$, where 
\begin{align*}
	K_{\phi_i}   & = V_{\phi_i}^{-1} + \frac{1}{\sigma_i^2}\sum_{t=2}^{T}(h_{i,t-1}-\mu_i)^2\\
	\hat{\phi}_h & = K_{\phi_i}^{-1}\left[V_{\phi_i}^{-1}\phi_{0,i} + \frac{1}{\sigma_i^2}\sum_{t=2}^{T}(h_{i,t-1}-\mu_i) (h_{i,t}-\mu_i) \right].
\end{align*}
Then, given the current draw $\phi_i$, a proposal $\phi_i^*$ is accepted with probability $\min(1,g(\phi_i^*)/g(\phi_i))$; 
otherwise the Markov chain stays at the current state $\phi_i$.

\textbf{Step 7}: lastly, sampling $\vkappa=(\kappa_1,\kappa_2)'$ can be done similarly as in other stochastic volatility models. More specifically, define the index set $S_{\kappa_1}$ that collects all the indexes $(i,j)$ such that $\alpha_{i,j}$, the $j$-th element of $\valpha_i$, is a coefficient associated with an own lag and let $S_{\kappa_2}$ denote the set that collects all the indexes $(i,j)$ such that $\alpha_{i,j}$ is a coefficient associated with a lag of other variables. Then, given the  priors $\kappa_j\sim \distn{G}(c_{j,1},c_{j,2}), j=1,2 $, and the prior covariance matrix of $\valpha_i$, we have
\begin{align*}
	(\kappa_{1} \gvn \valpha) & \sim 	\distn{GIG}\left(c_{1,1}-\frac{np}{2}, 2c_{1,2}, \sum_{(i,j)\in S_{\kappa_1}}\frac{(\alpha_{i,j}-\alpha_{0,i,j})^2}{C_{i,j}}\right) \\
	(\kappa_2 \gvn \valpha) & \sim \distn{GIG}\left(c_{2,1}-\frac{(n-1)np}{2}, 2c_{2,2},\sum_{(i,j)\in S_{\kappa_2}} \frac{(\alpha_{i,j}-\alpha_{0,i,j})^2}{C_{i,j}}\right), 	
\end{align*}
where $\alpha_{0,i,j}$ is the $j$-th element of the prior mean vector $\valpha_{0,i}$ and $C_{i,j}$ is a constant determined by the Minnesota prior. 

Next, we turn to the computation of the marginal likelihood of VAR-FSV. Here the marginal likelihood estimator has two parts: the conditional Monte Carlo where we integrate out the VAR coefficients and the latent factors; and the adaptive importance sampling that biases the joint distribution of $\bh, \mathbf{l}, \vmu, \vphi$ and $\vkappa$. In what follows, we first derive an analytical expression of the conditional Monte Carlo estimator $\Em[p(\by \gvn \valpha,\bl, \mathbf{f},\bh,\vkappa) \gvn \bl,\bh,\vkappa] = p(\by\gvn\bl,\bh,\vkappa)$. To that end, write the VAR-FSV model as
\[
	\by = \bX\valpha + (\mathbf{I}_T\otimes\bL)\bbf, \quad \bu^y\sim\distn{N}(\mathbf{0},\bD),
\]
where $\bX$ an appropriately defined covariate matrix consisting of intercepts and lagged values and $\bbf \sim\distn{N}(\mathbf{0},\bG)$. Hence, the distribution of $\by$ marginal of the factors $\bbf$ is $(\by\gvn \valpha, \bl,\bh) \sim \distn{N}(\bX\valpha,\bS_{y})$,
where $\bS_{y} = \bD + (\mathbf{I}_T\otimes\bL)\bG(\mathbf{I}_T\otimes\bL')$. Next, following a similar calculation as in the case for VAR-SV, one can show that 
\begin{align*}
	p(\by \gvn \bl, \bh,\vkappa) & = \int p(\by\gvn \valpha, \bl,\bh)p(\valpha\gvn \vkappa) \di\valpha \\
	& = (2\pi)^{-\frac{Tn}{2}}|\bS_y|^{-\frac{1}{2}}|\bV_{\valpha}|^{-\frac{1}{2}}
	|\bK_{\valpha}|^{-\frac{1}{2}}\e^{-\frac{1}{2}\left(\by'\bS_y^{-1}\by
		+ \valpha_{0}'\bV_{\valpha}^{-1}\valpha_{0}
		- \hat{\valpha}'\bK_{\valpha}\hat{\valpha}\right)},
\end{align*}
where $\bK_{\valpha} = \bV_{\valpha}^{-1} + \bX'\bS_y^{-1}\bX$ and $\hat{\valpha} = \bK_{\valpha}^{-1}(\bV_{\valpha}^{-1}\valpha_0 + \bX'\bS_y^{-1}\by).$ 

We can now write the marginal likelihood as:
\begin{align*}
	p(\by) & = \int p(\by\gvn\bl,\bh,\vkappa) p(\bl)	p(\vmu)p(\vphi)p(\vsigma^2)p(\vkappa) \prod_{j=1}^{n+r}p(\bh_{j,\bigcdot}\gvn\mu_j,\phi_j,\sigma_j^2)\di(\bl,\bh,\vmu,\vphi,\vsigma^2,\vkappa) \\
	& = \int p(\by\gvn\bl,\bh,\vkappa) p(\bl)	p(\vmu)p(\vphi)p(\vkappa) \prod_{j=1}^{n+r}p(\bh_{j,\bigcdot}\gvn\mu_j,\phi_j)\di(\bl,\bh,\vmu,\vphi,\vkappa),
\end{align*}
where $p(\bh_{i,\bigcdot}\gvn\mu_i,\phi_i) = \int p(\bh_{i,\bigcdot}\gvn\mu_i,\phi_i,\sigma^2_i)p(\sigma^2_i)\di\sigma^2_i$ is available analytically as given in~\eqref{eq:den_hmarg}.

To combine the conditional Monte Carlo with the importance sampling approach described in Section~\ref{ss:IS}, we consider the parametric family
\[
	\mathcal{F} = \left\{f_{\distn{N}}(\mathbf{\bl}; \hat{\bl},	\hat{\bK}_{\bl}^{-1})
	\prod_{j=1}^{n+r}f_{\distn{N}}(\bh_{j,\bigcdot}; \hat{\bh}_{j,\bigcdot}, \hat{\bK}_{\bh_{j,\bigcdot}}^{-1})
	f_{\distn{N}}(\mu_j; \hat{\mu}_j, \hat{K}_{\mu_j}^{-1})
	f_{\distn{N}}(\phi_j; \hat{\phi}_j, \hat{K}_{\phi_j}^{-1}) \prod_{k=1}^{2}
	f_{\distn{G}}(\kappa_k; \hat{\nu}_{\kappa_k}, \hat{S}_{\kappa_k})	
	\right\}.
\]
The parameters of the importance sampling densities are chosen by solving the maximization problem in~\eqref{eq:maxMC}. In particular, all the $T$-variate Gaussian densities are obtained using the procedure described in Section~\ref{ss:CSV}. Other low dimensional importance sampling densities can be obtained following \citet{CE15}.

\newpage
\subsection*{Appendix B: Data}

The dataset is sourced from the Federal Reserve Bank of St. Louis and covers the quarters from 1959:Q1 to 2019:Q4. Table~\ref{tab:var} lists the 30 quarterly variables and describes how they are transformed. For example, $\Delta \log $ is used to denote
the first difference in the logs, i.e., $\Delta \log x = \log x_t - \log x_{t-1}$.

\begin{table}[H]
\centering
\caption{Description of variables used in the empirical application.} \label{tab:var}
\resizebox{\textwidth}{!}{\begin{tabular}{llccc}
\hline\hline
Variable 	&	Transformation	&	$n=7$	&	$n=15$	&	$n=30$	\\ \hline
Real Gross Domestic Product 	&	400$\Delta \log$	&	x	&	x	&	x	\\
\rowcolor{lightgray}
Personal Consumption Expenditures	&	400$\Delta \log$	&		&	x	&	x	\\
Real personal consumption expenditures: Durable goods	&	400$\Delta \log$	&		&		&	x	\\
\rowcolor{lightgray}
Real Disposable Personal Income	&	400$\Delta \log$	&		&		&	x	\\
Industrial Production Index	&	400$\Delta \log$	&	x	&	x	&	x	\\
\rowcolor{lightgray}
Industrial Production: Final Products	&	400$\Delta \log$	&		&		&	x	\\
All Employees: Total nonfarm	&	400$\Delta \log$	&		&	x	&	x	\\
\rowcolor{lightgray}
All Employees: Manufacturing	&	400$\Delta \log$	&		&		&	x	\\
Civilian Employment	&	400$\Delta \log$	&		&	x	&	x	\\
\rowcolor{lightgray}
Civilian Labor Force Participation Rate	&	no trans. &		&		&	x	\\
Civilian Unemployment Rate	&	no trans.	&	x	&	x	&	x	\\
\rowcolor{lightgray}
Nonfarm Business Section: Hours of All Persons	&	400$\Delta \log$	&		&		&	x	\\
Housing Starts: Total	&	400$\Delta \log$	&		&	x	&	x	\\
\rowcolor{lightgray}
New Private Housing Units Authorized by Building Permits	&	400$\Delta \log$	&		&		&	x	\\
Personal Consumption Expenditures: Chain-type Price index	&	400$\Delta \log$	&		&	x	&	x	\\
\rowcolor{lightgray}
Gross Domestic Product: Chain-type Price index	&	400$\Delta \log$	&		&		&	x	\\
Consumer Price Index for All Urban Consumers: All Items	&	400$\Delta \log$	&	x	&	x	&	x	\\
\rowcolor{lightgray}
Producer Price Index for All commodities	&	400$\Delta \log$	&		&		&	x	\\
Real Average Hourly Earnings of Production and & & & & \\
Nonsupervisory Employees: Manufacturing	&	400$\Delta \log$	&	x	&	x	&	x	\\
\rowcolor{lightgray}
Nonfarm Business Section: Real Output Per Hour of All Persons	&	400$\Delta \log$	&		&		&	x	\\
Effective Federal Funds Rate	&	no trans.	&	x	&	x	&	x	\\
\rowcolor{lightgray}
3-Month Treasury Bill: Secondary Market Rate	&	no trans.	&		&		&	x	\\
1-Year Treasury Constant Maturity Rate	&	no trans.	&		&		&	x	\\
\rowcolor{lightgray}
10-Year Treasury Constant Maturity Rate	&	no trans.	&	x	&	x	&	x	\\
Moody's Seasoned Baa Corporate Bond Yield Relative to  & & & & \\
Yield on 10-Year Treasury Constant Maturity	&	no trans.	&		&	x	&	x	\\
\rowcolor{lightgray}
3-Month Commercial Paper Minus 3-Month Treasury Bill	&	no trans.	&		&		&	x	\\
Real M1 Money Stock	&	400$\Delta \log$	&		&	x	&	x	\\
\rowcolor{lightgray}
Real M2 Money Stock	&	400$\Delta \log$	&		&		&	x	\\
Total Reserves of Depository Institutions	&	$\Delta^2 \log$	&		&		&	x	\\
\rowcolor{lightgray}
S\&P's Common Stock Price Index : Composite	&	400$\Delta \log$	&		&	x	&	x	\\ \hline\hline
\end{tabular}}
\end{table}

\newpage

\subsection*{Appendix C: Additional Simulation Results}

This appendix provides various additional simulation results to further assess the performance of the proposed marginal likelihood estimators. 

The first set of results assesses whether one can distinguish the three stochastic volatility models---the common stochastic volatility model (VAR-CSV), the Cholesky stochastic volatility model (VAR-SV) and the factor stochastic volatility model (VAR-FSV)---in higher dimensions. More specifically, we repeat the Monte Carlo exercise in Section 4.1 of the main text, but here each VAR has $n=20$ variables instead of $n=10$. 

We first generate 100 datasets from VAR-CSV as described as Section 4.1, and for each dataset, we compute the log marginal likelihoods of VAR-CSV and VAR-FSV relative to that of VAR-CSV. The results reported in Figure~{fig:MC1-CSV-n20} show that for all datasets, the proposed marginal likelihood estimators select the correct DGP. In fact, compared to Figure~\ref{fig:MC1-CSV} based on VARs with $n=10$ variables, here the differences in the marginal likelihood values are larger. This suggests that it becomes easier to discriminate among the stochastic volatility models as the number of variables increases.

\begin{figure}[H]
  \centering
  \includegraphics[width=.8\textwidth]{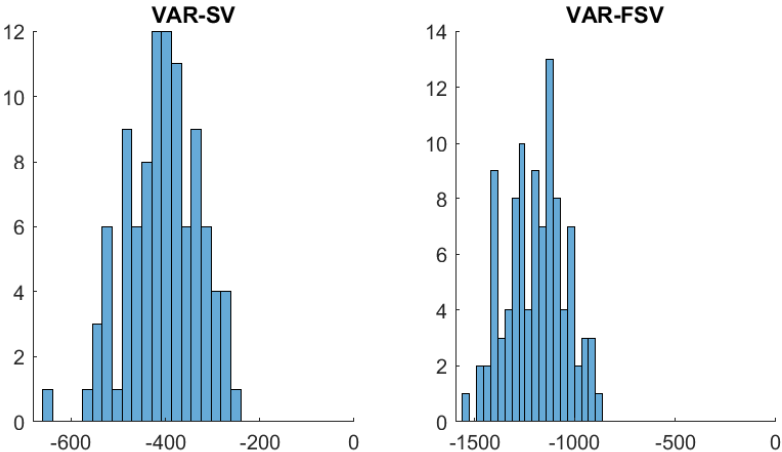}
  \caption{Histograms of log marginal likelihoods under VAR-SV (left panel) and VAR-FSV (right panel) relative to the true model (VAR-CSV) for $n=20$. A negative value indicates that the correct model is favored.}
   \label{fig:MC1-CSV-n20}
\end{figure}

\begin{figure}[H]
  \centering
  \includegraphics[width=.8\textwidth]{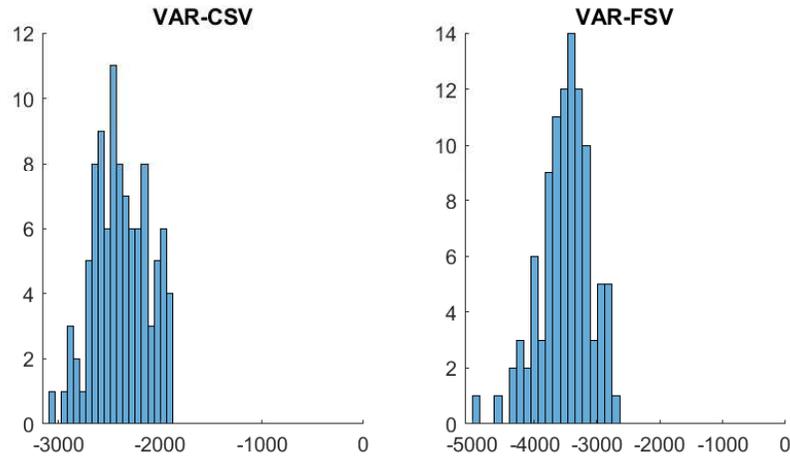}
  \caption{Histograms of log marginal likelihoods under VAR-CSV (left panel) and VAR-FSV (right panel) relative to the true model (VAR-SV) for $n=20$. A negative value indicates that the correct model is favored.}
   \label{fig:MC1-SV-n20}
\end{figure}

\begin{figure}[H]
  \centering
  \includegraphics[width=.8\textwidth]{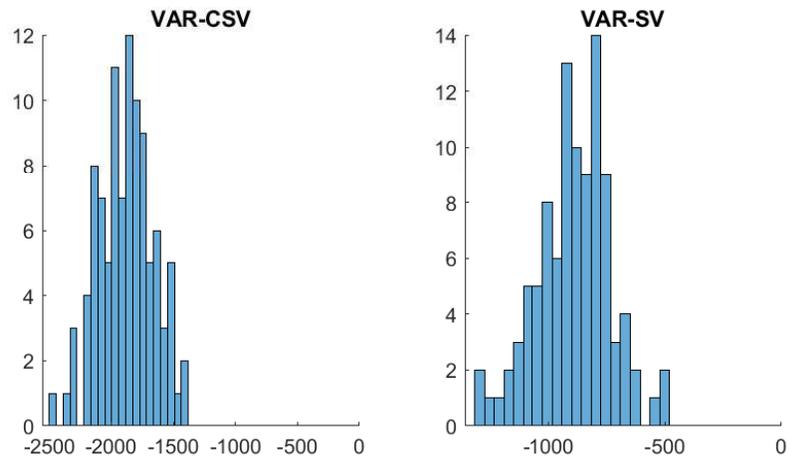}
  \caption{Histograms of log marginal likelihoods under VAR-CSV (left panel) and VAR-SV (right panel) relative to the true model (VAR-FSV) for $n=20$. A negative value indicates that the correct model is favored.}
   \label{fig:MC1-FSV-n20}
\end{figure}

Figures~\ref{fig:MC1-SV-n20} and \ref{fig:MC1-FSV-n20} present results based on datasets generated from VAR-SV and VAR-FSV, respectively. Once again, these model comparison results show that, for all datasets, the correct model is overwhelmingly favored compared to the other two stochastic volatility specifications. 

The next set of results assesses how the proposed marginal likelihood estimators perform when the magnitude of the shocks in the VAR equation increases. More specifically, we use the same settings as described in Section 4.1 of the main text, but here we increase the magnitude of the error covariance matrix in the VAR equation by a factor of 10. In particular, for VAR-CSV, the error covariance matrix $\vSigma$ is generated from the inverse-Wistart distribution $\distn{IW}(n+5, 7\times \mathbf{I}_n + 3\times\mathbf{1}_n\mathbf{1}_n')$; for VAR-SV and VAR-FSV, we increase the values of $\mu_i, i=1,\ldots, n$, i.e., the means of the log-volatility in the VAR equation, from $-1$ to 1.3. This difference of 2.3 in natural log corresponds to approximately a 10-fold increase in volatility. 

\begin{figure}[H]
  \centering
  \includegraphics[width=.8\textwidth]{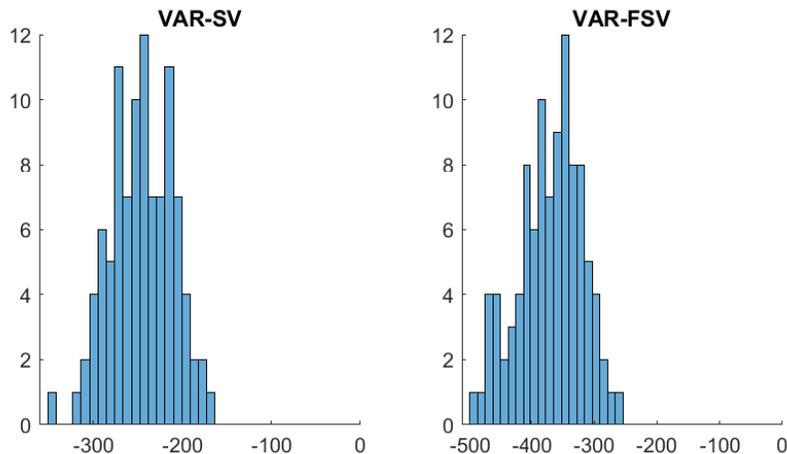}
  \caption{Histograms of log marginal likelihoods under VAR-SV (left panel) and VAR-FSV (right panel) relative to the true model (VAR-CSV) with a noisier VAR equation. A negative value indicates that the correct model is favored.}
   \label{fig:MC1-CSV-noisy}
\end{figure}

\begin{figure}[H]
  \centering
  \includegraphics[width=.8\textwidth]{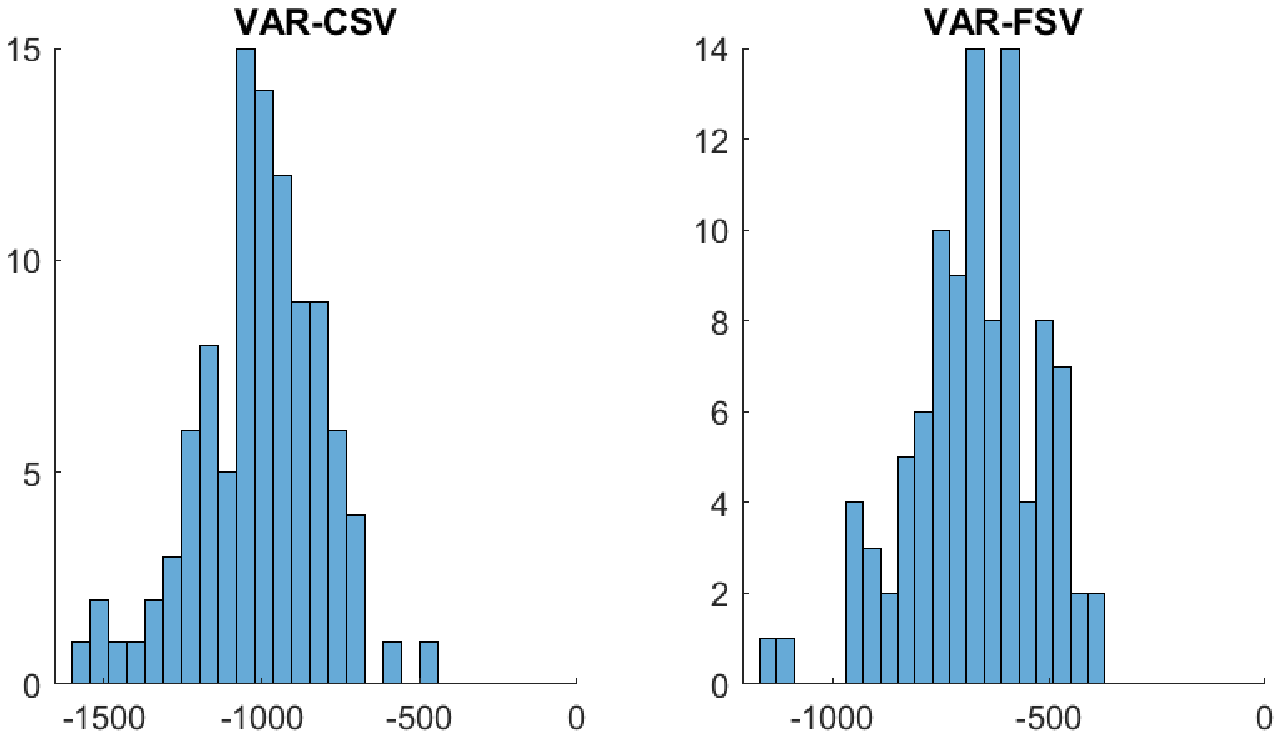}
  \caption{Histograms of log marginal likelihoods under VAR-CSV (left panel) and VAR-FSV (right panel) relative to the true model (VAR-SV) with a noisier VAR equation. A negative value indicates that the correct model is favored.}
   \label{fig:MC1-SV-noisy}
\end{figure}

\begin{figure}[H]
  \centering
  \includegraphics[width=.8\textwidth]{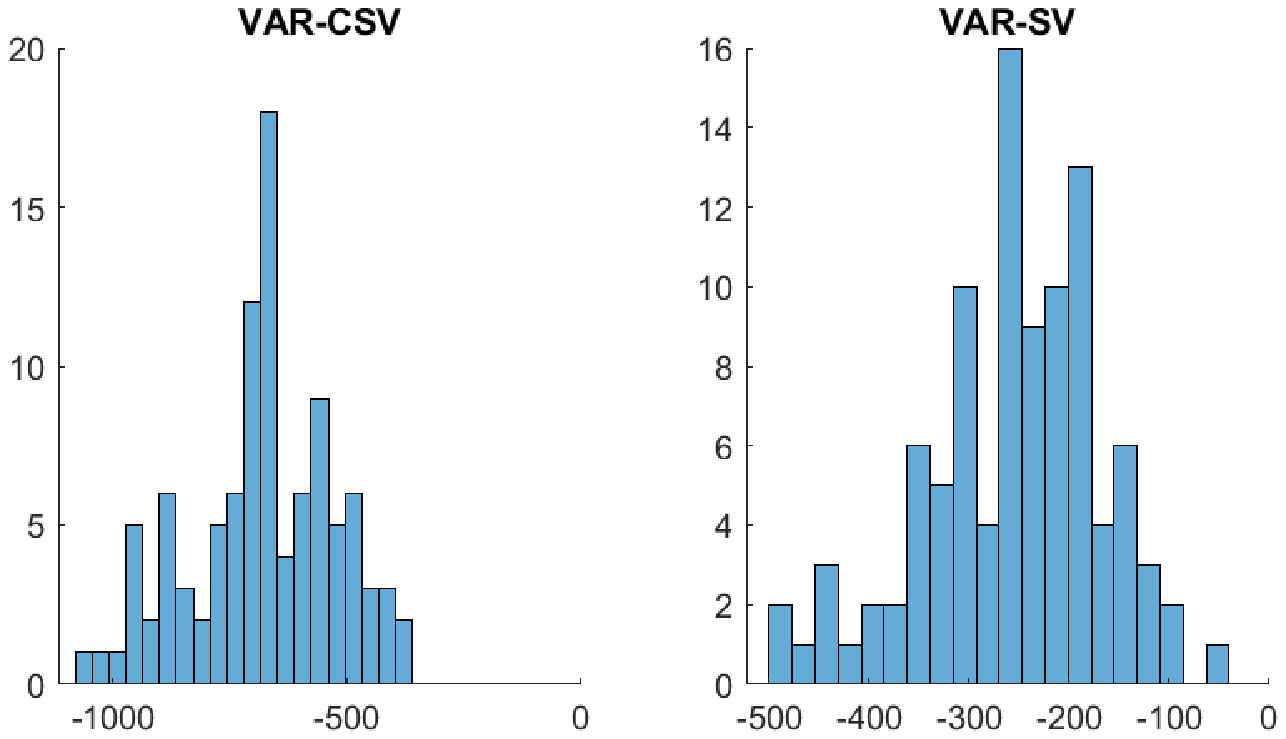}
  \caption{Histograms of log marginal likelihoods under VAR-CSV (left panel) and VAR-SV (right panel) relative to the true model (VAR-FSV) with a noisier VAR equation. A negative value indicates that the correct model is favored.}
   \label{fig:MC1-FSV-noisy}
\end{figure}

The model comparison results are reported in Figures~\ref{fig:MC1-CSV-noisy}-\ref{fig:MC1-FSV-noisy}. These results from the noisier DGPs are very similar to those reported in Section 4.1. In particular, for all datasets, the correct model is overwhelmingly favored compared to the other two stochastic volatility models. One minor difference in the two sets of results is that when the data is generated from VAR-FSV, it becomes slightly harder to distinguish between VAR-SV and VAR-FSV (reflected in the smaller differences in marginal likelihood values). 

\newpage

\singlespace

\ifx\undefined\BySame
\newcommand{\BySame}{\leavevmode\rule[.5ex]{3em}{.5pt}\ }
\fi
\ifx\undefined\textsc
\newcommand{\textsc}[1]{{\sc #1}}
\newcommand{\emph}[1]{{\em #1\/}}
\let\tmpsmall\small
\renewcommand{\small}{\tmpsmall\sc}
\fi

\end{document}